\documentstyle[aasms4,12pt]{article}
\begin{document}
\title{ The motion of stars near the Galactic center:
A comparison of the black hole and fermion ball scenarios}
\author{Faustin Munyaneza and Raoul D. Viollier}
\affil{Institute of Theoretical Physics and Astrophysics\\
 Department of Physics, University of Cape Town\\
 Private Bag, Rondebosch 7701, South Africa\\
 fmunyaneza@hotmail.com, viollier@physci.uct.ac.za
}

\begin{abstract}
After a discussion of the properties of degenerate fermion balls, we
analyze the orbits of the stars S0-1 and S0-2, which have the smallest
projected distances to Sgr A$^{*}$, in the supermassive black hole as well
as in the fermion ball scenarios of the Galactic center. It is shown that
both scenarios are consistent with the data, as measured during the last six
years by Genzel et al. and Ghez et al. The free parameters of the projected
orbit of a star are the unknown components of its velocity $v_{z}$
and distance $z$ to Sgr A$^{*}$ in 1995.4, with the $z$-axis being in the line
of sight. We show, in the case of S0-1 and S0-2, that the $z-v_{z}$ phase-
space, which fits the data, is much larger for the fermion ball than for the
black hole scenario. Future measurements of the positions or radial velocities
of S0-1 and S0-2 could reduce this allowed phase-space and eventually rule out
one of the currently acceptable scenarios. This may shed some light into the
nature of the supermassive compact dark object, or dark matter in general at
the center of our Galaxy.
\end{abstract}

\keywords{black hole physics-celestial mechanics, stellar dynamics
dark matter - elementary particles - Galaxy: center}

\section{Introduction}
There is strong evidence for the existence of a supermassive compact dark
object near the enigmatic radio source Sagittarius A$^{*}$ (Sgr A$^{*}$) which
is located
at or close to the dynamical center of the Galaxy (Rogers et al. 1994; Genzel et
al. 1997; Lo et al. 1998; Ghez et al. 1998). Stars observed in the 2.2 $\mu$m
infrared K-band at projected distances $\gtrsim$ 5 mpc from Sgr A$^{*}$, and
moving with projected velocities
$\lesssim$ 1400 km~s$^{-1}$, indicate that a mass
of (2.6 $\pm$ 0.2) $\times$ 10$^{6} M_{\odot}$ must be concentrated within a
radius $\sim$ 15 mpc from Sgr A$^{*}$ (Haller et al. 1996; Eckart and Genzel
1996, 1997; Genzel \& Townes 1987; Genzel et al. 1994, 1996, 1999, 2000;
Ghez et al. 1998, 2000). VLBA radio
interferometry measurements at 7 mm wavelength constrain the size of
the radio wave emitting region of Sgr A$^{*}$ to $\lesssim$ 1 AU in E-W
direction and $\sim$ 3.6 AU in N-S direction
(Rogers et al. 1994, Bower and Backer 1998, Krichbaum et al. 1994, Lo et al.
1998), and the proper motion of Sgr A$^{*}$ relative to the quasar background
to
$\lesssim$ 20 km~s$^{-1}$
(Baker 1996; Reid et al. 1999; Baker and Sramek 1999). As the fast moving
stars of the central cluster interact gravitationally with Sgr A$^{*}$, the
proper motion of the radio source cannot remain as small as it is now
for $\sim$ 200 kyr unless
Sgr A$^{*}$ is attached to some mass $\gtrsim$ 10$^{3} M_{\odot}$.
In spite of these well-known stringent facts, the enigmatic radio source Sgr
A$^{*}$, as well as the supermassive compact dark object that is perhaps
associated with it, are still two of the most challenging
mysteries of modern astrophysics.

It is currently believed that the enigmatic radio source Sgr A$^{*}$ coincides
in position with a supermassive black hole (BH) of (2.6 $\pm$ 0.2) $\times$
10$^{6} M_{\odot}$ at the dynamical center of the Galaxy. Although standard
thin
accretion disk theory fails to explain the peculiar low
luminosity $\lesssim$ 10$^{37}$ erg s$^{-1}$ of the Galactic center
(Goldwurm et al. 1994), many
models have been developed that describe the spectrum of Sgr A$^{*}$ fairly
well, based on the assumption that it is a BH. The models proposed for the
radio emission, range from quasi-spherical inflows (Melia 1994; Narayan and
Mahadevan 1995; Narayan et al. 1998; Mahadevan 1998)
to a jet-like outflow (Falcke, Mannheim and
Biermann 1993; Falcke and Biermann 1996; Falcke and Biermann 1999).
Yet, as some of these models appear to contradict each other, not all
of them can represent the whole truth.
We also
note that the Galactic center is a weak source of diffuse emission in the 2-10
keV energy range and in the lines of several ions (Sunyaev et al. 1993; Koyama
et al. 1996; Sidoli and Mereghetti 1999). Thus, apart from earthbound
VLBA radio interferometers, space missions such as the European
Multi-Mirror satellite (XMM) and Chandra X-ray satellite,
may eventually
provide conclusive evidence for the nature of Sgr A$^{*}$ and the supermassive
compact dark
object at the Galactic center. In fact, the Chandra X-ray satellite
 has
recently detected a point source at the location of Sgr A$^{*}$ (Baganoff et
al. 1999) with a luminosity two times smaller than the upper limit set by the
ROSAT satellite some years ago (Predehl and Tr\"umper 1994). For more
detailed recent reviews on the Galactic center we refer to Morris and Serabyn
1996, Genzel and Eckart 1999, Kormendy and Ho 2000, and Yusef-Zadeh et al.
2000.

Supermassive compact dark objects have also been inferred at the centers of many
other
galaxies, such as M87 (Ford et al. 1994; Harms et al. 1994; Macchetto et al.
1997) and NGC 4258 (Greenhill et al. 1995; Myoshi et al. 1995).
For recent reviews we refer to Richstone et al. 1998,
Ho and Kormendy 2000, and Kormendy 2000. In fact,
perhaps with the exception of dwarf galaxies, all galaxies may harbor such
supermassive compact dark objects at their centers. However, only a small
fraction of these
show strong radio emission similar to that of the enigmatic radio source Sgr
A$^{*}$ at the center of our Galaxy. For instance M31 does not have such a
strong compact radio source, although the supermassive compact dark object at
the center of M31 has
a much larger mass ($\sim$ 3 $\times$ 10$^{7} M_{\odot}$) than that of our
Galaxy
(Dressler and Richstone 1988; Kormendy
1988). It seems, therefore, prudent not to take for granted that the
enigmatic radio source Sgr A$^{*}$ and the supermassive compact dark object at
the center of our Galaxy are necessarily one and the same
object.

An unambiguous proof for the existence of a BH requires the observation of
stars moving at relativistic velocities near the event horizon. However, in
the case of our Galaxy, the stars S0-1 and S0-2, that are presumably closest
to the suspected BH, reach
projected velocities $\lesssim$ 1400 km~s$^{-1}$.
Assuming a radial velocity of $v_{z}$ = 0, this corresponds to the escape
velocity at a distance $\gtrsim$ 5 $\times$ 10$^{4}$ Schwarzschild radii
from the
BH. Thus any dark object,
having a mass $\sim$ 2.6 $\times$ 10$^{6} M_{\odot}$ and a
radius $\lesssim$ 5 $\times$ 10$^{4}$ Schwarzschild radii,
would fit the current
data on the proper motion of the stars of the central cluster as well as the
BH scenario. One of the reasons why the BH scenario of Sgr A$^{*}$ is so
popular, is that the only baryonic alternative to a BH that we can imagine, is
a cluster of dark stars (e.g. brown dwarfs, old white dwarfs, neutron stars,
etc.), having a total mass of $\sim$ 2.6 $\times$ 10$^{6} M_{\odot}$
concentrated within a radius of $\sim$ 15 mpc. However, such a star cluster
would disintegrate through gravitational ejection of stars on a
time scale $\lesssim$
100 Myr, which is much too short to explain why this object still seems to be
around today $\sim$ 10 Gyr after its likely formation together with the
Galaxy (Sanders 1992; Haller et al. 1996; Maoz 1995, 1998).
Nevertheless, in order to test the validity of the BH hypothesis
meaningfully, we definitely need an alternative and consistent finite size
model of the supermassive compact dark objects at the galactic centers.

\section{ The case for degenerate fermion balls}
It is well known that our Galactic halo is dominated by dark matter,
the bulk part of which must be nonbaryonic (Alcock 2000). Numerical
simulations show that dark matter in the form of a gas of weakly interacting
massive particles, will eventually produce a high-density spike at the
center of the Galaxy (Navarro et al. 1997; Gondolo \& Silk 1999). It is
therefore conceivable that
the supermassive compact dark object at the center of our Galaxy is made of
the same dark matter that dominates the Galactic halo at large. In fact, some
years
ago, we suggested that the supermassive compact dark object at the Galactic
center may be a gravitationally stable ball of weakly interacting fermions in
which the degeneracy pressure balances the gravitational attraction of the
massive fermions (Viollier et al. 1992, 1993; Viollier 1994; Tsiklauri \&
Viollier 1996; Bili\'{c}, Munyaneza \& Viollier 1999).
Such degenerate fermion balls (FBs) could
have been formed in the early universe during a first-order gravitational
phase transition (Bili\'{c} \& Viollier 1997, 1998, 1999a,b). A further
formation mechanism of FBs that is based on gravitational ejection of
degenerate matter has recently
been discussed in Bili\'{c} et al. 2000.

There are three main reasons why it is worthwhile to study such degenerate FBs
as an alternative to BHs at the center of the galaxies, in particular our own:
\begin{itemize}
\item[(i)] Introducing a weakly interacting fermion
in the $\sim$ 13 keV/$c^{2}$ to $\sim$ 17 keV/$c^{2}$ mass range, one can
explain the full range of the masses and radii of the supermassive compact
dark objects, that have been observed so far at the galactic centers, in terms
of degenerate FBs
with
masses ranging from 10$^{6}$ to 10$^{9.5} M_{\odot}$ (Kormendy and Richstone
1995; Richstone et al. 1998). The maximal mass allowed for a FB composed of
degenerate fermions of a given mass $m_{f}$ and degeneracy factor $g_{f}$ is
the Oppenheimer-Volkoff (OV) limit $M_{OV}$ = 0.54195 $M_{P \ell}^{3}$ $m_{f}^{-
2}$ $g_{f}^{- \frac{1}{2}}$ = 2.7821 $\times$ 10$^{9} M_{\odot}$
(15 keV/$m_{f}c^{2}$)$^{2}$(2/$g_{f}$)$^{\frac{1}{2}}$,
where $M_{P \ell} = (\hbar c/G)^{\frac{1}{2}}$ is the Planck mass (Bili\'{c},
Munyaneza \& Viollier 1999). It is tempting to identify the mass of the most
massive compact dark object ever observed at a center of a galaxy
(Kormendy \& Ho 2000), e.g. that
of the center of M87, with the OV-limit, i.e. $M_{OV}$ = (3.2 $\pm$ 0.9)
$\times$ 10$^{9} M_{\odot}$ (Macchetto et al. 1997). This requires a fermion
mass of 12.4 keV/$c^{2}$ $\lesssim$ $m_{f}$ $\lesssim$ 16.5 keV/$c^{2}$ for
$g_{f}$ = 2,
or 10.4 keV/$c^{2}$ $\lesssim$ $m_{f}$ $\lesssim$ 13.9 keV/$c^{2}$ for $g_{f}$ =
4.
For $M_{OV}$ = 3.2 $\times$ 10$^{9} M_{\odot}$ such a
relativistic FB would have a radius of $R_{OV}$ = 4.45 $R_{OV}^{s}$ = 1.36
mpc, where $R_{OV}^{s}$ is the Schwarzschild radius of the mass $M_{OV}$.
It would thus be virtually indistinguishable from
a BH, as the radius of the last stable orbit around a BH is 3 $R_{OV}^{s}$ =
0.92 mpc
anyway. The situation is quite different for a nonrelativistic FB of mass
$M$ = (2.6 $\pm$ 0.2) $\times$ 10$^{6} M_{\odot}$, which for the upper limit
of the allowed fermion mass ranges, $m_{f}$ = 16.5 keV/$c^{2}$ for $g_{f}$ = 2,
or $m_{f}$ = 13.9 keV/$c^{2}$ for $g_{f}$ = 4, would have a radius bound by
16.7 mpc $\lesssim$ $R$ $\lesssim$ 17.6 mpc, corresponding to $\sim$ 7 $\times$
10$^{4}$ Schwarzschild radii, as the FB radius scales nonrelativistically like
$R \propto m_{f}^{-8/3}$ $g_{f}^{- 2/3}$ $M^{-1/3}$. Such an object is far from
being a black hole: its escape velocity from the center is
$\sim$ 1,700 km~s$^{-1}$. As the fermions interact only weakly with the
baryons,
baryonic stars could also move inside a FB without experiencing noticeable
friction with the fermions
(Tsiklauri and
Viollier 1998a,b; Munyaneza, Tsiklauri and Viollier, 1998, 1999).
Since the potential within $\sim$ 10 mpc from the center is rather shallow,
star formation in this region will be less inhibited
by tidal forces than in the BH case.
\item[(ii)] A FB with mass $M$ = (2.6 $\pm$ 0.2) $\times$ 10$^{6}
M_{\odot}$ and radius $R$ $\lesssim$ 18.4 mpc is consistent with the current
data on the proper motion of the stars in the central cluster around Sgr
A$^{*}$. This implies lower limits for the fermion masses of $m_{f}$ $\gtrsim$
15.9 keV/$c^{2}$ for
$g_{f}$ = 2 and $m_{f}$ $\gtrsim$ 13.4 keV/$c^{2}$ for $g_{f}$ = 4, which partly
overlap with the fermion mass ranges derived for M87. By increasing
the
fermion mass, one can interpolate between the FB and the BH scenarios.
However, for fermion masses $m_{f}$ $\gtrsim$ 16.5 keV/$c^{2}$, for
$g_{f}$ = 2 and $m_{f} \gtrsim$ 13.9 keV/$c^{2}$ for $g_{f}$ = 4, the
interpretation of some of the most massive compact dark objects in terms of
degenerate FBs is no longer possible. It is quite remarkable that we can
describe the two extreme cases, the supermassive compact dark object at the
center of M87 and that of our Galaxy, in terms of self-gravitating degenerate
FBs using a single fermion mass. This surprising fact is a consequence of the
 equation of state of degenerate fermionic matter; this would
not be the case for degenerate bosonic matter. Indeed, for a
supermassive object consisting of nonrelativistic self-gravitating degenerate
bosons, mass and radius would scale, for a constant boson mass, as $R \propto
M^{-1}$, rather than $R \propto M^{-1/3}$, as for a supermassive object
consisting of nonrelativistic self-gravitating degenerate fermions, for a
constant fermion mass. The ratio of the radii of the supermassive objects with
10$^{6.5} M_{\odot}$ and 10$^{9.5} M_{\odot}$ would be 10$^{3}$ in the boson
case, instead of 10 as in the fermion case. Thus it would not be possible
to fit mass and radius of both the supermassive compact dark object at the
center of M87 and that of our Galaxy, in the boson case. We therefore conclude
that,
if we want to describe all the supermassive compact dark objects in terms of
self-gravitating degenerate particles of the same kind and mass, these objects
cannot be composed of bosons, they must consist of fermions.
\item[(iii)] The FB scenario provides a natural cut-off of the emitted
radiation at infrared frequencies $\gtrsim$ 10$^{13}$ GHz, as is
actually observed in the spectrum of the Galactic center (Bili\'{c}, Tsiklauri
and Viollier 1998; Tsiklauri and Viollier 1999; Munyaneza and Viollier 1999).
This is because matter, e.g. in the form of stars, gas, dust or
dark matter, etc.
falling from infinity at rest towards the FB, cannot
acquire velocities larger than the escape velocity from the center of the FB,
i.e. $\sim$ 1,700 km~s$^{-1}$. Consequently, there is also a natural cut-off of
the high-frequency tail of the radiation emitted by the accreted baryonic
matter. This is quite a robust prediction of the FB scenario, because it is
virtually independent of the details of the accretion model. In a thin disk
accretion model, the radiation at the observed cut-off is emitted at
distances $\sim$ 10 mpc from the center of the FB. This is also the region,
where the gravitational potential becomes nearly harmonic due to the finite
size of the FB. The fermion masses required for a cut-off at the observed
frequency $\sim$ 10$^{13}$ GHz
depend somewhat on the accretion rate and the inclination angle of the disk assumed,
but
$m_{f}$ $\lesssim$ 20 keV/$c^{2}$ for
$g_{f}$ = 2 or $m_{f}$ $\lesssim$ 17 keV/$c^{2}$ for $g_{f}$ = 4
seem to be reasonable conservative upper limits (Tsiklauri and Viollier 1999,
Munyaneza and Viollier 1999).
\end{itemize}
Summarizing the preceding arguments (i) to (iii), we can constrain the
allowed fermion masses for the supermassive compact dark objects in our Galaxy
to 15.9 keV/$c^{2}$ $\lesssim$ $m_{f}$ $\lesssim$ 16.5 keV/$c^{2}$ for
$g_{f}$ = 2 or 13.4 keV/$c^{2}$ $\lesssim$ $m_{f}$ $\lesssim$ 13.9
keV/$c^{2}$ for $g_{f}$ = 4, where the lower limits are determined from
the proper motion of stars in
the central cluster of our Galaxy, while the upper limits arise from the
supermassive compact dark object at the center of M87.
This fermion mass
range is
also consistent with the infrared cut-off of the radiation emitted by the
accreted baryonic matter at the Galactic center. Of course, one of the major
challenges will be to accommodate, within the FB scenario, the properties
of Sgr A$^{*}$ which is perhaps peculiar to our galaxy.

We now would like to identify a suitable candidate for the postulated weakly
interacting
fermion. This particle should have been either already observed, or its
existence
should have been at least predicted
in recent elementary particle theories.
 The required fermion cannot be the gaugino-like neutralino,
i.e. a linear combination of the bino, wino and the two higgsinos,
as its mass is expected to be in the $\sim$ 30 GeV/$c^{2}$ to
$\sim$ 150 GeV/$c^{2}$ range (Roszkowski 2001).
It cannot be a standard
neutrino either (however, see Giudice et al. 2000),
as this would violate the cosmological bound on neutrino mass
and, more seriously, it would contradict the Superkamiokande data
(Fukuda et al. 2000). However, the
required fermion could be the sterile neutrino that has been recently
suggested as a cold dark matter candidate in the mass range between $\sim$ 1
keV/$c^{2}$ to
$\sim$ 10 keV/$c^{2}$ (Shi and Fuller 1999; Chun \& Kim 1999; Tupper et al.
2000), although one would have to stretch the mass range a little bit and
worry about the (possibly too rapid) radiative decay into a standard neutrino.
This sterile neutrino is resonantly produced with a cold spectrum and near
closure density,
if the initial lepton asymmetry is $\sim$ 10$^{-3}$.
Alternatively, it could be either the gravitino, postulated in supergravity
theories with a mass in the $\sim$ 1 keV/$c^{2}$
to $\sim$ 100 GeV/$c^{2}$ range (Lyth 1999), or the axino, with a mass
in the range between $\sim$ 10 keV/$c^{2}$ and $\sim$ 100 keV/$c^{2}$, as
predicted by the
supersymmetric extensions of the Peccei-Quinn solution to the strong
CP-problem (Goto \& Yamaguchi 1992). In this scenario, the axino mass arises
quite naturally as a radiative correction
in
a model with a no-scale superpotential. In summary, there are at least three
promising candidates which have been recently predicted for completely
different reasons
in elementary particle theories. One of these particles could play the role of
the weakly interacting
fermion required for the supermassive compact dark objects at the centers of
the galaxies and for cold or warm dark matter at large, if its mass is in the
range between
$\sim$ 13 keV/$c^{2}$ and $\sim$ 17 keV/$c^{2}$ and its contribution to the
critical density is $\Omega_{f} \sim$ 0.3.

\section{Outline of the paper}
The purpose of this paper is to compare the predictions of the BH and FB
scenarios of the Galactic center, for the stars with the smallest projected
distances to Sgr A$^{*}$, based on the measurements of their positions during
the last six years (Ghez et al. 2000). The projected orbits of three stars,
S0-1 (S1), S0-2 (S2) and S0-4 (S4), show deviations from uniform motion on a
straight line during the last six years, and they thus may contain nontrivial
information about the potential.
We do not rely on the accelerations
determined directly from the data by Ghez et al. 2000, as this was
done in the constant acceleration approximation which we think is not
reliable. Indeed, the Newtonian predictions for the acceleration vary
substantially, both in
magnitude and direction, during the six years of observation.
In view of this fact, we prefer to work with the raw data directly, trying
to fit
the projected positions in right ascension (RA) and declination
of the stars in the BH and FB scenarios. For our analysis we have selected only
two stars,
S0-1 and S0-2, because their projected distances from SgrA$^{*}$ in 1995.53,
4.42 mpc and 5.83 mpc, respectively, make it most likely that
these could be orbiting within a FB of radius $\sim$ 18 mpc. We thus may
in principle distinguish between the BH and FB scenarios for these two stars.
The third star, S0-4, that had in 1995.53 a projected distance
of 13.15 mpc from Sgr A$^{*}$, and was moving away from Sgr A$^{*}$ at a
projected
velocity of $\sim$ 990 km~s$^{-1}$, is now definitely
outside a FB with a radius $\sim$ 18 mpc. One would thus not be
able to distinguish the two scenarios for a large part of the
 orbit of S0-4.

In the following, we perform a detailed analysis of the orbits of the stars S0-1
and S0-2, based
on the Ghez et al. 1998 and 2000 data, including the error bars of the
measurements, and varying the unknown components of the position and velocity
vectors of the stars in 1995.4, $z$ and $v_{z}$.
For simplicity, we assume throughout this paper that the supermassive compact
dark object has a
mass of 2.6 $\times$ 10$^{6} M_{\odot}$, and is centered at the position of
Sgr A$^{*}$ which is taken to be at a distance of 8 kpc from the sun. In fact,
because of the small proper motion $\lesssim$ 20 km~s$^{-1}$ of Sgr A$^{*}$,
there are strong dynamical reasons to assume in the BH scenario, that Sgr
A$^{*}$ and the supermassive BH are at the same position, while in the FB
scenario, Sgr A$^{*}$ and the FB could be off-center by a few mpc without
affecting
the results. We do not vary the mass of the supermassive compact dark object,
as the calculations are not very sensitive to this parameter, as long as the
mass
is within the range of the error bar inferred from the
statistical data on the proper motion of the stars in the central
cluster (Ghez et al. 1998).

This paper is organized as follows: In section 4, we present the main equations
for the description of the supermassive compact dark object as a FB,
as well as the formalism for the description of the dynamics of the stars
in the gravitational field of a FB
or a BH.
We then investigate, in section 5,
the dynamics of
S0-1 and S0-2, based on the Ghez et al. 2000 data,
and conclude with a summary and outlook in section 6.

\section{The dynamics of the stars near the Galactic center}
As the stars near the Galactic center have projected velocities $\lesssim$
1,400 km~s$^{-1}$, one may very well describe their dynamics
in terms of Newtonian mechanics for both the BH and the FB scenarios.
Similarly, fermions of mass $m_{f} \sim$ 13 keV/$c^{2}$ to $\sim$ 17
keV/$c^{2}$,
which are condensed
in a degenerate FB of (2.6 $\pm$ 0.2) $\times$ 10$^{6} M_{\odot}$, are
nonrelativistic,
since their local Fermi velocity is certainly smaller than the escape
velocity of $\sim$ 1,700 km~s$^{-1}$ from the center of the FB. The
fermions will, therefore, obey the
equation of hydrostatic equilibrium, the Poisson equation and the
nonrelativistic equation of state of degenerate fermionic matter
\begin{equation}
P_{f} = K n_{f}^{5/3}
\end{equation}
with
\begin{equation}
K = \frac{\hbar^{2}}{5 m_{f}} \; \left( \frac{6 \pi^{2}}{g_{f}} \right)^{2/3}
 \; \; .
\end{equation}
Here, $P_{f}$ and $n_{f}$, denote the local pressure and particle number density
of the fermions, respectively. FBs have been discussed extensively in a number
of
papers (e.g. Viollier 1994; Bili\'{c}, Munyaneza and Viollier 1999;
Tsiklauri and Viollier 1999). Here we merely quote the equations
that we need further below, in order to make this paper self-contained.
The gravitational potential of a degenerate FB
is given by
\begin{eqnarray}
\Phi (r) =
\left\{
\begin{array}{l}
 \displaystyle{\frac{GM_{\odot}}{a} \; \left( v'(x_{0}) - \frac{v(x)}{x}
 \right) \; \; , \; \; x \leq x_{0} } \\ [.5cm]
 \displaystyle{- \frac{GM}{ax} \hspace{3.35cm} , \; \; x >  x_{0} } \; \; ,\\
\end{array} \right.
\end{eqnarray}
where $a$ is an appropriate unit of length
\begin{equation}
a \; = \; \left( \frac{3 \pi \hbar^{3}}{4 \sqrt{2}\; m_{f}^{4}\; g_{f}\; G^{3/2}
\; M_{\odot}^{1/2}} \right)^{2/3} \; = \; 0.94393\;\mbox{pc} \;
\left( \frac{15 \; \mbox{keV}}{m_{f} c^{2}} \right)^{8/3} \; g_{f}^{- 2/3} \;
\; ,
\end{equation}
$r = ax$ is the distance from the center of the FB and $R = ax_{0}$ the
radius of the FB.
The dimensionless quantity $v(x)$, that is related to the gravitational
potential $\Phi (r)$ through eq.(3),
obeys the Lan\'{e}-Emden  differential
equation
\begin{equation}
\frac{d^{2} v}{d x^{2}} \; = \; - \; \frac{v^{3/2}}{x^{1/2}} \; \; ,
\end{equation}
with polytropic index $n = 3/2$. For a pure FB without a
gravitational point source at the center, the boundary conditions at the
center and the surface of the FB are $v(0) = v(x_{0}) = 0$.
All the relevant quantities
of the FB can be expressed in terms of $v$ and $x$, e.g. the matter
density as
\begin{equation}
\rho \; = \; \frac{\sqrt{2}}{3} \; \frac{m_{f}^{4}\; g_{f}}{\pi^{2}
\hbar^{3}} \; \left( \frac{GM_{\odot}}{a} \right)^{3/2} \; \left( \frac{v}{x}
\right)^{3/2} \; \; ,
\end{equation}
where $m_{f}$ and $g_{f}$ are the mass and the spin degeneracy factor of
the fermions and antifermions, respectively,
i.e. $g_{f}$ = 2 for Majorana and $g_{f}$ = 4 for Dirac fermions and
antifermions. Based on eqs.(5) and (6), the mass enclosed within a radius $r$
in a FB is given by
\begin{equation}
M(r) \; = \; \int_{0}^{r} \; 4 \pi \; \rho \; r^{2} \; dr \; = \;
- \; M_{\odot} \; \left(v'(x) x - v(x) \right) \; \; ,
\end{equation}
and the total mass of the FB by
\begin{equation}
M \; = \; M(R) \; = \; - \; M_{\odot} \; v'(x_{0}) \; x_{0} \; \; .
\end{equation}
From eq.(5), one can derive a scaling relation for the mass and radius of a
nonrelativistic FB, i.e.
\begin{eqnarray}
M R^{3} &=& x_{0}|v'(x_{0})|x_{0}^{3} \; a^{3} \; M_{\odot} \; = \;
\frac{91.869 \; \hbar^{6}}{G^{3} m_{f}^{8}} \; \left( \frac{2}{g_{f}}
\right)^{2} \nonumber \\
&=& 27.836 \; M_{\odot} \; \left( \frac{15 \; \mbox{keV}}{m_{f} c^{2}}
\right)^{8}
\;
\left( \frac{2}{g_{f}} \right)^{2} \; (\mbox{pc})^{3} \; \; .
\end{eqnarray}
Here $v(x)$ is the solution of eq.(5) with $v(0)$ = 0 and $v'(0)$ = 1,
yielding $v(x_{0})$ = 0 again at $x_{0}$ = 3.65375, and $v'(x_{0})$ = --
0.742813. The precise index of the power law of the scaling relationship (9)
depends on the
polytropic index of the equation of state (1). As the mass of the FB approaches
the OV
limit, this scaling law is no longer valid, because the degenerate fermion gas
has
to be described by the correct relativistic equation of state and
Einstein's equations for the gravitational field and hydrostatic equilibrium
(Bili\'{c}, Munyaneza \& Viollier 1999).

We now turn to the description of the dynamics of the stars near the Galactic
center. The mass of the BH and FB is taken to be $M$ = 2.6 $\times$ 10$^{6}
M_{\odot}$. In order to emphasize the differences between the FB and the BH
scenarios, we choose the fermion masses $m_{f}$ = 15.92 keV/$c^{2}$ for
$g_{f}$ = 2 or $m_{f}$ = 13.39 keV/$c^{2}$ for $g_{f}$ = 4. These are the
minimal fermion masses consistent with the mass distribution inferred from
the statistics of proper motions of the stars in the central cluster (Munyaneza,
Tsiklauri and Viollier, 1999; Ghez et al. 1998). The dynamics of the stars in
the gravitational field of the supermassive compact dark object can be
calculated solving Newton's equations of motion
\begin{eqnarray}
\ddot{\vec{r}} \; = \; - \; \frac{GM(r)}{r^{3}} \; \vec{r} \; \; ,
\end{eqnarray}
taking into account the position and velocity vectors at e.g. $t_{0}$ =
1995.4 yr, i.e. $\vec{r}(t_{0}) \equiv (x,y,z)$ and $\dot{\vec{r}}(t_{0})
\equiv (v_{x}, v_{y}, v_{z})$. For the FB scenario, $M(r)$ is given by eq.(7),
while in the
BH case it is replaced by $M$ of eq.(8). The $x$-axis is chosen in the direction
opposite to the right ascension (RA), the $y$-axis in the direction of the
declination, and the $z$-axis points towards the sun. The BH and the center of
the FB are assumed to be at the position of Sgr A$^{*}$ which is also the
origin of the coordinate system at an assumed distance of 8 kpc from the sun.

\section{Analysis of the orbits of S0-1 and S0-2}
In 1995.4, the projected positions and velocities of S0-1 reported by
Ghez et al. 1998,
were $x = - 0.107''$, $y = 0.039''$, $v_{x}$ = (470 $\pm$ 130) km~s$^{-1}$ and
$v_{y}$ = (-1330 $\pm$ 140) km~s$^{-1}$. We now investigate how the
projected orbits,
calculated using eq.(10), are affected by (i) the error bars of $v_{x}$ and
$v_{y}$ of S0-1 measured in 1995.4, (ii) the lack of knowledge of $z$ of S0-1
in
1995.4, (iii) the lack of information on $v_{z}$ of S0-1 in 1995.4. We then
compare the results with the S0-1 data recently reported by Ghez et
al. 2000.

Fig.1 shows the RA of S0-1 as a function of time, taking into account
the error bars of $v_{x}$ and
$v_{y}$ and choosing $z = v_{z}$ = 0 in 1995.4.
The top panel represents the RA of S0-1 in the BH scenario, while the bottom
panel illustrates the same quantities in the FB case. From Fig.1 we conclude
that, for $z = v_{z}$ = 0 in 1995.4, the error bars of
$v_{x}$ and $v_{y}$ of 1995.4 do not allow for a fit of the new Ghez et al.
2000 data of S0-1 in the
BH scenario, whereas the data are described quite easily within the error
bars in the FB case. In Fig.2,
the declination is plotted as a function of time for the same values of $v_{x}$,
$v_{y}$, $z$ and $v_{z}$ in 1995.4. We arrive at the same conclusion
as in Fig.1: For $z = v_{z}$ = 0 the error bars of $v_{x}$ and $v_{y}$ allow
for a fit of the data in the FB scenario only.

As a next step, the sensitivity of the orbits to the $z$-
coordinate of S0-1 in 1995.4 is investigated. To this end, we restrict ourselves
to bound orbits of S0-1 only.
The conserved total energy of the star S0-1 is given by
\begin{equation}
E \; = \; \frac{1}{2} m \; \dot{\vec{r}}^{2}
\; + \; m \Phi (r) \; \; ,
\end{equation}
where the unknown star mass can be chosen as $m$ = 1 without loss of
generality.
S0-1 is unlikely to have a total energy $E > 0$, because,
in the absence of swing-by events caused by stars of the central cluster, S0-1
will have to fall in with a velocity that is inconsistent with the velocity
dispersion of the stars at infinity.
The condition $E \leq$ 0 thus yields upper limits,
$|v_{z}| \leq |v_{z}^{\infty}|$ and $|z| \leq |z_{\infty}|$,
which depend on
$v_{x}$ and $v_{y}$ as can be seen from Table~1.
In this context, it is worthwhile to note that, at a radius equal to the
projected distance of S0-1 to Sgr A$^{*}$ in
1995.4, the escape
velocity from a BH is $2,250~{\rm km \ s^{-1}}$, while
that from a FB is $1,613~{\rm km \ s^{-1}}$.
The escape velocity from the center of the FB is 1,672 km s$^{-1}$.

Fig.~3 presents the sensitivity of the RA of S0-1 with respect to the
$z$-coordinate
in both the BH and FB scenarios.
In the case of a BH, the RA depends strongly on the value of $z$ in
1995.4, while the
$z$-dependence in the FB scenario is rather weak.
In both the top and bottom panels,
$v_{x}=340 \ {\rm km \ s^{-1}}$,
$v_{y}=-1190 \ {\rm km  s^{-1}}$ and $v_{z}=0$ has been assumed while
$z$ is varied, all in 1995.4 values.
In the BH scenario, none of the other input values for $v_{x}$ and $v_{y}$
would
fit the new Ghez et al. 2000 data if we restrict ourselves to
bound orbits.
In the FB case, the input values $v_{x}=340~{\rm km \ s^{-1}}$ and
$v_{y}=-1470~{\rm km \  s^{-1}}$ describe the Ghez et al. 2000 data as well as
$v_{x}=340~{\rm km \ s^{-1}}$ and
$v_{y}=-1190~{\rm  km \ s^{-1}}$.
Looking at Fig. 3, we thus conclude that the RA data of S0-1 are well
fitted with
$|z| \approx 0.25''$ in the BH scenario, and
with $|z| \lesssim 0.1''$
in the FB case.
The dependence of the declination on $z$ is shown in Fig.4.
In order to describe
the Ghez et al. 2000 data for the declination of S0-1, we require
$0.25'' \lesssim |z| \lesssim |z_{\infty}| = 0.359''$
for the BH scenario, while in the FB case,
the declination can be fitted with
$|z| \lesssim 0.359''$.

Fig.5 represents an investigation of how the RA of S0-1 is affected by $v_{z}$
of
1995.4.
In this graph we have chosen $v_{x}$ = 340 km~s$^{-1}$,
$v_{y}$ = - 1190 km~s$^{-1}$
and $z$ = 0 in 1995.4.
Increasing
$|v_{z}|$ up to
its maximal value $|v_{z}^{\infty}|$ = 1879 km~s$^{-1}$ does not help fitting
the RA data of S0-1
in the BH scenario.
In the FB case, the orbits are rather insensitive to
$|v_{z}|$. Thus all $|v_{z}| \; \lesssim \;
|v_{z}^{\infty}|$ = 1036 km~s$^{-1}$
fit the RA data of S0-1 quite well.
The weak dependence of the RA on $|v_{z}|$ in the FB scenario is due to the
harmonic
oscillator like shape of the FB potential at small distances, where the
Newtonian
equations of motion nearly decouple in Cartesian coordinates.
Fig.6 exhibits the declination as a function of time for
various values of $|v_{z}|$, keeping $v_{x}$, $v_{y}$ and $z$ in 1995.4
as in Fig.5. The top panel of Fig.6 shows that increasing $|v_{z}|$ from zero up
to its
maximal value $|v_{z}^{\infty}|$ = 1879 km~s$^{-1}$ barely helps fitting
the data in the BH scenario.
In the FB scenario, the declination may be described by
 $|v_{z}| \lesssim$ 900 km~s$^{-1}$.

Summarizing the results of Figs.3-6, we can plot in Fig.7 the $z-v_{z}$
phase-space of 1995.4
that fits
the data. The small range of acceptable
$|z|$ and $|v_{z}|$ values in the BH scenario (solid vertical line)
reflects the fact that the
orbits of S0-1 depend strongly
on these two parameters.
Conversely, the weak dependence of the orbits on $|z|$ and $|v_{z}|$
in the FB case is the reason for the much larger $z-v_{z}$ phase-space that is
fitting the Ghez et al. 2000
data of S0-1, as shown by the dashed box. The dashed and solid curves
describe the $E$ = 0 (just bound) orbit in the FB and BH scenarios,
respectively.

Fig.8 shows some typical projected orbits of S0-1
in the BH and FB scenarios.
The Ghez et al. 2000 data of S0-1 may be fitted in both
scenarios with appropriate choices of $v_{x}$, $v_{y}$, $z$ and $v_{z}$
in 1995.4. The inclination angles of the orbit's plane
$\theta = \arccos \left( L_{z}/|\vec{L}|\right)$, with $\vec{L} = m \vec{r}
\times \dot{\vec{r}}$, are shown next to
the orbits.
The minimal
inclination angle that describes the data in the BH
case is $\theta = 70\arcdeg$, while in the FB scenario it is $\theta$ =
0$^{o}$.
In the BH case, the minimal
and maximal distances from Sgr A$^{*}$
are $r_{min} = 0.25''$ and $r_{max} = 0.77''$, respectively, for the orbit
with $z = 0.25''$ and $v_{z}$ = 0 which has a period of $T \approx $ 161 yr.
The orbits with $z = 0.25''$ and $v_{z}$ = 400 km s$^{-1}$
or $z = 0.25''$ and $v_{z}$ = 700 km~s$^{-1}$ have periods of $T \approx$ 268 yr
or
$T \approx$ 3291 yr, respectively.
In the FB scenario, the open orbit with $z = 0.1''$ and $v_{z}=0$ has a
``period''
of $T \approx$ 77 yr with $r_{min} = 0.13''$ and $r_{max} = 0.56''$.
The open orbits with $z = 0.1''$ and $v_{z}$ = 400 km~s$^{-1}$
or $z$ = 0.1'' and $v_{z}$ = 900 km~s$^{-1}$ have ``periods'' of
$T \approx$ 100 yr or $T \approx$
1436 yr, respectively.

Fig.9 is a prediction of $|v_{z}|$ as a function of time
for both the BH and FB scenarios and
various acceptable input parameters (see Fig.7).
This shows that the radial velocity
$|v_{z}|$ of S0-1, if measured in a few years time, could serve to distinguish
between the BH and the FB scenarios.
For the BH case, we predict by the year 2005 a radial velocity $|v_{z}| \;
\lesssim$ 900 km~s$^{-1}$, while in
the FB scenario the radial velocity
will be $|v_{z}| \; \lesssim$ 500 km~s$^{-1}$.
Radial velocities $|v_{z}| \; \gtrsim$ 1000 km~s$^{-1}$ before 2010 would
be excluded in both the BH and FB scenarios.

We now repeat the analysis in the case
of the star S0-2.
The $x$- and $y$-components of the position and velocity vectors
of S0-2
at $t_{0}$ = 1995.4 yr are $\vec{r} (t_{0})$ = ($0.007'',0.151''$) and
$\dot{\vec{r}} (t_{0})$ = (-290 $\pm$ 110, -500 $\pm$ 50) km~s$^{-1}$,
respectively (Ghez et al. 1998).
Restricting ourselves to bound orbits,
the Ghez et al. 2000 data of S0-2 can be fitted in the
BH scenario, with $v_{x}$ = -290 km~s$^{-1}$, $v_{y}$ = -500 km~s$^{-1}$,
$|z| \approx 0.25''$ and $|v_{z}| \lesssim$
1280 km~s$^{-1}$ in 1995.4.
In the FB scenario, the allowed $z-v_{z}$ phase-space is
$|z| \; \lesssim$ 0.1'' and $|v_{z}| \lesssim$
1000 km~s$^{-1}$ for the same values of $v_{x}$ and $v_{y}$ in 1995.4.
The range of acceptable values of $|z|$ and $|v_{z}|$ is shown in Fig.10,
where the solid and dashed curves denote the limits on
$|z|$ and $|v_{z}|$ for $E$ = 0 (just bound) orbits in the BH and FB cases,
respectively. Here again the $z-v_{z}$ phase-space turns out to be much
larger in the FB (dashed box) than in the BH scenario (vertical solid
line).

Fig.11 exhibits some typical projected orbits of S0-2
corresponding
to the acceptable $z-v_{z}$ phase-space of Fig.10 in both the BH and FB
scenarios.
Next to the orbits the inclination angles of the orbit's plane
are shown.
In the BH scenario, the lower limit of the inclination angle is $\theta
= 74\arcdeg$, for the
orbit (1) with $z = 0.25''$ and $v_{z} = 0$.
The minimal and maximal distances from the center are
$r_{min} =0.045''$ and $r_{max} = 0.307''$, respectively.
The period of this closed orbit is minimal with $T \approx$ 34 yr.
However, this value disagrees with the minimal period $T \approx$
16 yr quoted by Ghez et al. 2000. The orbit (2) with $z = 0.25''$ and
$v_{z}$ = 500 km~s$^{-1}$ or
$z$ = 0.25'' and $v_{z}$ =1000 km~s$^{-1}$ have periods $T \approx$ 42 yr or
$T \approx$ 135 yr, respectively. In the FB scenario the minimal inclination
angle of the orbit's plane is $\theta = 0\arcdeg$. The open orbit (1) with
$z = 0.25''$  and $v_{z} = 0$ has minimal and maximal distances from the center
of
$r_{min} = 0.15''$ and $r_{max} = 0.31''$, respectively,
with a ``period'' of $T \approx$ 46 yr.
The open orbits with (1) $z = 0.25''$ and $v_{z}$ = 500 km~s$^{-1}$ or
(2) $z = 0.25''$ and $v_{z}$ = 1000~km~s$^{-1}$ have ``periods''
$T \approx$ 51 yr or
$T \approx$ 174 yr, respectively.
The minimal ``period''
$T \approx$ 37 yr
is obtained for an open orbit with $z = v_{z}$ = 0
and inclination angle $\theta = 0\arcdeg$.

In Fig. 12 the predictions for the radial velocity $|v_{z}|$
are plotted as a function
of time for various acceptable input values of $z$ and $v_{z}$ in
1995.4.
It is seen that a radial velocity of
$|v_{z}| \; \gtrsim$ 1000 km~s$^{-1}$
is excluded in the FB scenario.

\section{Summary and outlook}
In this paper, we have shown that the orbits of S0-1
and S0-2, observed by Ghez et al. 2000 during the last six years, are consistent
with either a BH or a FB of 2.6
$\times$ 10$^{6} M_{\odot}$ at the center of the Galaxy.
In order to fit the data in the BH scenario, S0-1
and S0-2 must have had in 1995.4 a $z$-coordinate of $|z| \approx 0.25''$
and radial velocities of $|v_{z}| \; \lesssim$ 750 km~s$^{-1}$ or
$|v_{z}| \; \lesssim$ 1280 km~s$^{-1}$ for S0-1 or S0-2, respectively.
In the BH scenario, the orbits of S0-1 and S0-2 strongly depend on $z$
in 1995.4. The new data of S0-1 and S0-2 can be
fitted in the FB scenario with $z = v_{z}$ = 0 in 1995.4. Due to the weak
dependence
of the orbits on $|z|$ and $|v_{z}|$ in the FB case,
$|z|  \lesssim 0.1''$ and $|v_{z}| \lesssim$ 900 km s$^{-1}$ for S0-1 or
$|z|  \lesssim 0.25''$ and
$|v_{z}| \; \lesssim$ 1000 km~s$^{-1}$ for S0-2 are
also consistent with the Ghez et al. 2000 data. As new
measurements become available, the acceptable $z-v_{z}$
phase-space of 1995.4 could be further constrained.

We have plotted some typical orbits of
S0-1 and S0-2 in both the BH and FB scenarios and have shown that the
minimal inclination angle of the orbit's plane is $\theta = 70\arcdeg$ as
in the BH case and $\theta = 0\arcdeg$
in the FB scenario.
We have established that by the year 2005, the measurement of the
radial velocities $|v_{z}|$ of
both S0-1 and S0-2 could discriminate between the two scenarios of the
supermassive compact dark object at the Galactic center. In concluding,
it is important to note again that, based on the Ghez et al. 2000 data of the
stars S0-1 and S0-2,
the FB scenario cannot be ruled out. On the
contrary, in view of the $z-v_{z}$ phase-space, that is much larger in FB
scenario than in the BH case, there is reason to
treat the FB scenario of the supermassive compact dark object at the center of
our Galaxy with the respect it deserves.

We now turn to the discussion of promising techniques for proving or
disproving the FB or BH scenarios of the Galactic center. These can be basically
divided into two classes:
\begin{itemize}
\item[(i)] probing the gravitational potential at distances $\lesssim 0.5''$
from the Galactic center, and
\item[(ii)] observing the decay or the annihilation of the fermions of the FB
into visible particles, at distances $\lesssim 0.5''$ from the Galactic
center.
\end{itemize}
In the first category, the most promising method is still
monitoring the proper motions and radial velocities of stars that are
located
at
projected distances $\lesssim 0.5''$ from Sgr A$^{*}$, as well as
interpreting these observations in terms of the FB or BH scenarios. However,
a further interesting possibility is the observation of gravitationally
lensed stars in the line of sight behind
the FB (Bili\'{c}, Nikoli\'{c} and
Viollier 2000), as a FB of $\sim$ 18 mpc radius is a much more
efficient gravitational lens than a BH of the same mass. In fact, a star
crossing the line of sight with a minimal distance of $\sim$ 0.2 mpc,
$\sim$ 200 pc behind the center of the FB, will
produce for a few years up to three distinct moving images within
or just outside the Einstein ring
radius of $\sim 0.13''$ or $\sim$ 5 mpc. Two of these star images will pop
out of nothing at some point inside the Einstein
ring, and they will first separate, then approach each
other again and finally annihilate each other at a different point
within the Einstein ring.
The third image moves around the Einstein ring while the projection of the
source crosses the ring area.
Unfortunately, the rate for such a remarkable
event, which would be the smoking gun for an extended supermassive object, is
estimated to be only $\sim$ 10$^{-4}$/yr.
In the BH case, the manifestations of gravitational lensing are less
spectacular with only two observable lensed moving images (Wardle and
Yusef-Zadeh 1992).
Another possible test of the gravitational potential
could be the spectrum of the radiation emitted by the accreted
baryonic matter,
once the
model dependence of the calculations can be controlled.

In the second category, the particle content of FBs could be proven e.g. through
the
radiative decay of the fermion (assumed here to be a sterile neutrino) into a
standard neutrino, i.e. $f
\rightarrow \nu \gamma$. If the lifetime for this decay is 1.4 $\times$
10$^{18}$ yr, the luminosity of the FB would be 4 $\times$ 10$^{34}$ erg/sec.
The signal would be a sharp X-ray line of $\sim$ 8 keV
for $g_{f}$ = 2 or $\sim$ 7 keV for $g_{f}$ = 4. The X-ray luminosity would be
tracing the fermion matter density. Of course the spatial resolution of the X-
ray telescope would have to be $\lesssim 0.5''$. A further possible test could
be the annihilation of two fermions into two or three photons, i.e. $f \bar{f}
\rightarrow \gamma \gamma$ or $f \bar{f} \rightarrow \gamma \gamma \gamma$.
However, the branching ratios of these two annihilation processes
with respect to a presumably dominant but unobservable
$f \bar{f} \rightarrow \nu' \bar{\nu'}$ channel
are most probably too
small to be observable. Nevertheless, the signal would be a sharp line
of $\sim$ 16 keV ($g_{f}$ = 2) of $\sim$ 14 keV ($g_{f}$ = 4) for
$f \bar{f} \rightarrow \gamma \gamma$, and a continuous spectrum with a maximum
at $\sim$ 10 keV ($g_{f}$ = 2) or $\sim$ 9 keV ($g_{f}$ = 4) for
$f \bar{f} \rightarrow \gamma \gamma \gamma$. For s-wave annihilation
the X-ray luminosity would
trace the square of the fermion matter density, while for p-wave
annihilation the concentration of the X-ray emission towards the center of
the FB would be even more pronounced.

\subsection*{Acknowledgements}
This work is supported in part by the Foundation for Fundamental Research (FFR)
grant number PHY 99-01241.
F.~Munyaneza gratefully acknowledges the financial support from the
British Department of Social Security (DSS).
We have enjoyed valuable discussions with
N. Bili\'{c}, T.
Koch, D. Tsiklauri and G.B. Tupper.

\newpage
\subsection*{Figure captions:}
Fig.1 : The RA of S0-1 as a function of time, taking into account the error bars
of the projected velocity in 1995.4 (Ghez et al. 98). The top panel is drawn
for the case of a black hole (BH), while the bottom panel corresponds to a
fermion ball (FB), both centered at Sgr A*. The labels of the graphs stand for
the
velocity components: (1) $v_{x}$ = 470 km~s$^{-1}$, $v_{y}$ = -1330
km~s$^{-1}$ (median values); (2) $v_{x}$ = 340 km
s$^{-1}$, $v_{y}$ = -1190 km~s$^{-1}$; (3) $v_{x}$ = 340 km~s$^{-1}$,
$v_{y}$ = -1470 km~s$^{-1}$;
(4) $v_{x}$ = 600 km~s$^{-1}$, $v_{y}$ = -1190 km~s$^{-1}$; (5) $v_{x}$ = 600
km~s$^{-1}$, $v_{y}$ = -1470 km~s$^{-1}$, all for 1995.4.
In this plot, $z$ and $v_{z}$ are taken to be zero in 1995.4.
One thus concludes that, within the error bars of the projected velocities of
1995.4, the data points recently reported by Ghez et al. 2000 can be fitted
in the FB scenario with $v_{z}=z$ = 0 (lines 2 and 3), while this is not
the case in the BH scenario. In this figure, the masses of both the FB
and the BH, are $M$ = 2.6 $\times$ 10$^{6}M_{\odot}$. The fermion mass is
taken to be $m_{f}c^{2}$ = 15.92
keV with a degeneracy factor of $g_{f}$ = 2, or $m_{f}c^{2}$ = 13.39 keV
with $g_{f}$~=~4.
Fig.2: This is an analysis similar to that of Fig.1 but for the
 declination of S0-1 as a function of time.
Assuming $v_{z}$ = $z$ = 0 in 1995.4, the
 data points can be represented within the error bars
 of $v_{x}$
 and $v_{y}$ in 1995.4 in the FB scenario only.
Fig. 3: Here we investigate the dependence of the RA of S0-1 on the
$z$-coordinate.
The data can be described with $v_{x}$ = 340
km s$^{-1}$, $v_{y}$ = -1190 km~s$^{-1}$, $v_{z}$ = 0 and
$|z| \approx 0.25''$
in
the BH case. In the FB scenario, the data can be fitted for
the same values of $v_{x}$, $v_{y}$, $v_{z}$ and $|z|
\; \lesssim 0.1'' $. We note that the RA strongly depends on $z$ in the BH
case, while in the FB scenario the dependence on $z$
is rather weak, as the equations of motion nearly decouple for small
$x,y$, and $z$. The limit value $|z_{\infty}| = 0.359''$ corresponds to a just
bound orbit with total energy $E=0$
and $v_{z}$ = 0 (see Table 1).
Fig.4 : The dependence of the declination of S0-1 on $|z|$ for $v_{z}$ = 0,
$v_{x}$ = 340 km~s$^{-1}$ and
$v_{y}$ = -1190 km~s$^{-1}$, all in 1995.4 values. The declination can be
fitted from $|z| \approx 0.25''$ to $|z_{\infty}| = 0.359''$ in the BH case.
In the FB scenario all values $|z|  \lesssim 0.359''$
fit the data.
Fig.5: Here we explore how the RA of S0-1 depends on $|v_{z}|$ while
keeping $z=0$, $v_{x}$ = 340 km~s$^{-1}$ and $v_{y}$ = -1190 km~s$^{-1}$,
taken in 1995.4. Increasing $|v_{z}|$ up to its maximal value
$|v_{z}^{\infty}|$ = 1879 km~s$^{-1}$, which corresponds to a
$E$ = 0 (just bound) orbit with $z$ = 0,
does not help fitting the data in the BH scenario. In the FB case,
there is a weak dependence on $v_{z}$, and all the values $|v_{z}| \; \lesssim$
1879 km~s$^{-1}$ describe the data.
Fig.6: The declination of S0-1 as a function of $|v_{z}|$ for $z=0$, $v_{x}$
= 340 km~s$^{-1}$ and $v_{y}$ = -1190
km s$^{-1}$, all in 1995.4 values. In the BH case, the data can only be
described for very large $|v_z|$, while in the FB scenario the data are fitted
for $|v_{z}| \; \lesssim$ 900 km~s$^{-1}$.
Fig.7: The $z-v_{z}$ phase-space of 1995.4 that fits the S0-1 data
with $v_{x}$ = 340 km~s$^{-1}$ and $v_{y}$ = - 1190 km~s$^{-1}$ in 1995.4.
In the BH case,the data require $|z| \approx 0.25''$ (vertical solid line), as
the orbits
strongly depend on $z$, and $|v_{z}| \; \lesssim$ 750 km~s$^{-1}$.
The large range of acceptable $|z|$ and $|v_{z}|$ values in the FB case,
denoted by the dashed rectangular box, is
due to the fact that the orbits depend weakly on these two parameters.
Values of $|z| \; \lesssim 0.1''$ and $|v_{z}| \; \lesssim$ 900 km~s$^{-1}$
are consistent with the measured positions of S0-1 in the FB scenario.
The solid and dashed curves correspond to the limits of possible values of
$|z|$ and $|v_{z}|$ for $E$ = 0 (just bound) orbits of S0-1 in the BH and FB
cases, respectively.
Fig.8: Examples of typical orbits of S0-1
for $v_{x}$ = 340 km~s$^{-1}$ and $v_{y}$ = - 1190 km~s$^{-1}$ in 1995.4.
The inclination angles of the
plane of the orbit are indicated next to the curves.
In the BH case, the orbit with $z = 0.25''$ and $v_{z}=0$ in 1995.4 has minimal
and maximal
distances from Sgr A* of $r_{min} = 0.25''$ and $r_{max}$ = 0.77'',
respectively, with a period of $T \approx$ 161 yr. The orbits with
$z = 0.25''$ and $v_{z}$ = 400 km~s$^{-1}$ or $z$ = 0.25'' and $v_{z}$ = 700
km~s$^{-1}$ in 1995.4 have
periods of $T \approx$ 268 yr or $T \approx$ 3291 yr, respectively.
In the FB scenario, the open orbit with $z = 0.1''$ and $v_{z}$ = 0 has a
''period'' of
$T \approx $ 77 yr with $r_{min} = 0.13''$ and $r_{max} = 0.56''$.
The open orbit with $z$ = 0.1'' and $v_{z}$ = 400 km~s$^{-1}$ in 1995.4 has a
''period'' $T \approx $ 100 yr,
while that with $z$ = 0.1'' and $v_{z}$ = 900 km~s$^{-1}$ has a
''period'' of $T \approx$ 1436 yr.
We note that the minimal inclination angle of the orbit's plane is
$\theta = 70\arcdeg$
in the BH case and $\theta = 0\arcdeg$ in the FB scenario.
Fig.9: The prediction for $|v_{z}|$ as a function of time for S0-1.
In the BH scenario with $|z| = 0.25''$ and $v_{z}$ = 0 in 1995.4, $|v_{z}|$
should be $\lesssim$ 900
km~s$^{-1}$ by the year 2005. The FB scenario predicts a
$|v_{z}|$ $\lesssim$ 500 km~s$^{-1}$ by the
year 2005, for $v_{z}$ = 0 and $|z|= 0.1''$ in 1995.4. Thus, once the radial
velocities are measured, this figure
could serve to distinguish between the BH and FB scenarios of Sgr A*.
Fig. 10: The $z-v_{z}$ phase-space of 1995.4 that fits the S0-2 data
with $v_{x}$ = - 290 km~s$^{-1}$ and $v_{y}$ = - 500 km~s$^{-1}$ in
1995.4. The solid and dashed curves denote the limits on $|z|$ and $|v_{z}|$
for $E$ = 0 (just bound) orbits in the BH and FB scenarios, respectively.
In the BH case, the vertical solid line at
$|z| \approx 0.25''$ up the BH curve stands for the region of the allowed
values of $|z|$ and $|v_{z}|$ for S0-2. In the FB scenario, the large
allowed $z-v_{z}$ phase-space
denoted by the dashed rectangular box, is due to the fact that the
open orbits depend rather weakly on these two parameters.
Fig.11: Examples of typical orbits of S0-2 for $v_{x}$ = - 290 km~s$^{-1}$
and
$v_{y}$ = - 500 km~s$^{-1}$ in 1995.4.
The labels of the orbits in both the upper (BH) and lower (FB) panels stand
for: (1) $z = 0.25''$, $v_{z}$ = 0; (2) $z$ = 0.25'', $v_{z}$ = 500 km~s$^{-
1}$; (3) $z = 0.25''$, $v_{z}$ = 1000 km~s$^{-1}$. The parameters of the label
(0) $z$ = 0, $v_{z}$ = 150 km~s$^{-1}$ fit the S0-2 data in the FB
scenario only.
The inclination angles
of the plane of the orbit are indicated next to the curves.
In the FB case, these values correspond to the maximal possible
$|z|$ and $|v_{z}|$ in 1995.4 for bound orbits. Bearing in mind that
$|z| \; \lesssim 0.25''$ in the FB case, the
inclination angle can take any value from
$\theta = 0\arcdeg$ to $\theta = 82\arcdeg$.
As an example, an orbit with $\theta = 29\arcdeg$
with the label (0) is shown in the lower panel.
However, in the BH scenario, the lower limit for the inclination angle
is $\theta = 74\arcdeg$. The closed orbit with label (1)
has $r_{min} = 0.045''$, $r_{max}= 0.307''$ and a period of
$T \approx$ 34 yr. The closed orbits with labels (2) and (3)
have periods of $T \approx$ 42 yr and $T \approx$ 135 yr, respectively.
The open orbit with label (1) has a ``period'' of $T \approx$ 46 yr with
$r_{min}$ = 0.15'' and $r_{max}$ = 0.31''. The open orbits with labels (2) and
(3) have ``periods'' of $T \approx$ 51 yr and $T \approx$ 174 yr,
respectively.
Fig.12: The prediction for $|v_{z}|$ as a function of time for S0-2.
This could serve to distinguish between the FB and the BH scenarios of
the supermassive dark object at the Galactic center. A radial velocity
$v_{z} \; \gtrsim$ 1000 km~s$^{-1}$
would rule out the FB scenario.
\newpage
\begin{table*}
\begin{center}
\begin{tabular}{|cc|cc|cc|} \hline
\multicolumn{2}{|c|}{}&  \multicolumn{2}{c|}{black hole}&
\multicolumn{2}{c|}{fermion ball}\\
 \hline
$v_{x} \  ({\rm km \ s^{-1}})$ & $v_{y} \ ({\rm km \ s^{-1}})$& $|z_{\infty}|$
\ (arcsec) & $|v_{z}^{\infty}| \ ({\rm km \ s^{-1}}$) &  $|z_{\infty}|$ \
(arcsec)&
$|v_{z}^{\infty}| \ ({\rm km \ s^{-1}}$)\\
 \hline
470 &  -1330&  0.27&   1753 &  0.24&   784 \\
340 & -1190&  0.359 & 1879&  0.359 & 1036 \\
340 & -1470&  0.226 & 1669& 0.165 & 572 \\
600 & -1190&  0.304& 1813& 0.295 & 910 \\
600 & -1470&  0.1985&  1594& 0.079&  287 \\
\hline
\end{tabular}
\end{center}
\caption{Maximal values of $|z|$ and $|v_{z}|$ in 1995.4 for $E$ = 0 (just
unbound) orbits of S0-1. \label{zvz}}
\end{table*}

\begin{thebibliography}{0-99}
\bibitem{} Alcock, C. 2000, Science, 287, 74
\bibitem{} Backer, D.C. 1996, in Unsolved problems in the Milky Way,
eds.L. Blitz and P. Teuben, Proc. of IAU Symp. No. 169
(Dordrecht: Kluwer)
\bibitem{} Backer, D. C.,  Sramek, R. A., 1999, ApJ, 524, 805
\bibitem{} Baganoff, F., et al. 1999, American Astronomical Society Meeting,
 195, 6201
\bibitem{} Bili\'c, N., Lindebaum, R. J., Tupper, G. B.,  Viollier, R. D.2000,
 astro-ph/0008230
\bibitem{} Bili\'c, N., Munyaneza, F., Viollier, R.D. 1999, Phys. Rev. D, 59,
 024003
\bibitem{} Bili\'c, N., Nikoli\'c, H.,  Viollier, R. D. 2000, ApJ, 537, 909
\bibitem{} Bili\'c, N., Tsiklauri, D.G.,  Viollier, R.D. 1998, Prog. Part.
Nucl.
 Phys., 40, 17
\bibitem{} Bili\'c, N.,  Viollier, R.D. 1997, Phys. Lett. B, 408,75
\bibitem{} Bili\'c, N.,  Viollier, R.D. 1998, Nucl. Phys. (Proc. Suppl.)
 B, 66,256
\bibitem{} Bili\'c, N.,  Viollier, R.D. 1999a, Gen. Relativ. Grav., 31,
 1105
\bibitem{} Bili\'c, N.,  Viollier, R.D. 1999b, Eur. Phys. J. C, 11,
 173
\bibitem{} Bower, G.C.  Backer, D.C. 1998, ApJ, 496, L97
\bibitem{} Chun, E. J.,  Kim, H. B. 1999, Phys. Rev. D, 60, 095006
\bibitem{} Dressler, A.  Richstone, D.O., 1988, ApJ, 324, 701
\bibitem{} Eckart, A.,  Genzel, R. 1996, Nature, 383,415
\bibitem{} Eckart, A.,  Genzel, R. 1997, MNRAS, 284, 576
\bibitem{} Falcke, H.,  Biermann, P. L. 1996, A\&A, 308, 321
\bibitem{} Falcke, H.,  Biermann, P. L. 1999, A\&A, 342, 49
\bibitem{} Falcke, H., Mannheim, K.,  Biermann, P. L. 1993, A\&A, 278, L1
\bibitem{} Ford, H.C., et al. 1994, ApJ, 435, L27
\bibitem{} Fukuda, S., et al. 2000, Phys. Rev. Lett., 85, 3999
\bibitem{} Genzel, R., Eckart, A., Ott, T.,  Eisenhauer, F. 1997,MNRAS,
291, 219
\bibitem{} Genzel, R., Eckart, A. 1999,
in Proc. The Central Parsecs of the Galaxy,
eds. Falcke, H., Cotera, A., Duschl, W., Melia, F., Rieke, M., ASP Conf.Series, Vol. 186
\bibitem{} Genzel, R., Hollenbach, D.J.,  Townes, C.H. 1994, Rep. Prog. Phys., 57,417
\bibitem{} Genzel, R., Pichon, C., Eckart, A., Gerhard, O. E. \&  Ott, T. 2000, MNRAS,317, 418
\bibitem{} Genzel, R., Thatte, N., Krabbe, A., Kroker, H., \& Tacconi-Garman, L.E. 1996, ApJ, 472, 153
\bibitem{} Genzel, R., \& Townes, C.H. 1987, ARA\&A, 25, 377
\bibitem{} Ghez, A.M., Klein, B.L., Morris, M., \& Becklin, E.E. 1998, ApJ, 509, 678
\bibitem{} Ghez, A. M., Morris, M.,Becklin, E. E., Tanner, A., \& Kremenek, T. 2000, Nature, 407, 351
\bibitem{} Giudice, G.F. et al. 2000, hep-ph/0012317.
\bibitem{} Goldwurm, A. et al. 1994, Nature, 371, 589
\bibitem{} Gondolo, P. \& Silk, J. 1999, Phys. Rev. Lett. 83, 1719
\bibitem{} Goto, T., \& Yamaguchi, M. 1992, Phys. Lett., B276, 123
\bibitem{} Greenhill, L.J., Jiang, D.R., Moran, J.M., Reid, M.J.,Lo, K.Y.,
 \& Claussen, M.J. 1995, ApJ, 440, 619
\bibitem{} Haller, J. W., Rieke, M. J., Rieke, G.H., Tamblyn, P., Close , L.,
\& Melia , F. 1996, ApJ, 456, 194
\bibitem{} Harms, R. J., et al. 1994, ApJ, 435, L35
\bibitem{} Ho, L. C., \& Kormendy, J., 2000, in Encyclopedia of Astronomy
 and Astrophysics, ed. Institute of Physics Publishing, astro-ph/0003267
\bibitem{} Kormendy, J. 1988, ApJ, 325, 128
\bibitem{} Kormendy, J. 2000, Nature, 407, 307
\bibitem{} Kormendy, J., \& Richstone, D. 1995, ARA\&A, 33, 581
\bibitem{} Kormendy, J., \& Ho, L.C., 2000, in Encyclopedia of Astronomy
 and Astrophysics, ed. Institute of Physics Publishing, astro-ph/0003268
\bibitem{} Koyama, K., et al., 1996, PASJ, 48, 249
\bibitem{} Krichbaum, T.P. et al. 1994, in Compact Extragalactic Radio Sources
(Proc. NRAO Workshop, Socorro, New Mexico), ed. J.A. Zensur and
 K.I. Kellermann (Greenbank: NRAO), 39
\bibitem{} Lo, K. Y., Shen, Z.-Q., Zhao, J.H., \& Ho, P. T. 1998, ApJ, 508, L61
\bibitem{} Lyth, D.H., hep-ph/9911257
\bibitem{} Macchetto, F., et al. 1997, ApJ, 489, 579
\bibitem{} Mahadevan, R. 1998, Nature, 394, 651
\bibitem{} Maoz, E. 1995, ApJ, 447, L91
\bibitem{} Maoz, E. 1998, ApJ, 494, L181
\bibitem{} Melia, F. 1994, ApJ, 426, 577
\bibitem{} Morris, M., \& Serabyn, E. 1996, ARA\&A, 34, 645
\bibitem{} Munyaneza, F., Tsiklauri, D., \& Viollier, R. D. 1998, ApJ, 509,
 L105
\bibitem{} Munyaneza, F., Tsiklauri, D., \& Viollier, R. D. 1999, ApJ, 526,
744
\bibitem{} Munyaneza, F.  \& Viollier, R.D., astro-ph/9907318
\bibitem{} Myoshi, M., Moran, J. M., Hernstein, J., Greenhill, L., Nakai,
N.,Diamond, P.,  Inoue, M. 1995, Nature, 373, 127
\bibitem{}Narayan, R., et al. 1998, ApJ, 492, 551
\bibitem{}Narayan, R., Yi, I., \& Mahadevan, R. 1995, Nature, 374,623
\bibitem{} Navarro, J.F., Frenk, C.S. \& White, S.D.M. 1997, ApJ, 490, 493
\bibitem{}Predehl, P., \& Tr\"umper, J. 1994, ApJ, 509, 678
\bibitem{}Reid, M. J., Readhead, A. C. S., Vermeulen, R. C., Treuhaft, R. N.
1999,ApJ, 524, 816
\bibitem{}Richstone, D., et al. 1998, Nature, 394, A14
\bibitem{}Rogers, A.E.E., et al. 1994, ApJ, 434, L59
\bibitem{} Roszkowski, L. 2001, hep-ph/0102327
\bibitem{}Sanders, R. H. 1992, Nature, 359, 131
\bibitem{}Shi, X., \& Fuller, G. M. 1999, Phys. Rev. Lett., 82, 2823
\bibitem{}Sidoli, L., \& Mereghetti, S. 1999, A\&A, 349, L49
\bibitem{}Sunyaev, V., et al. 1993, ApJ, 407, 603
\bibitem{}Tsiklauri, D., \& Viollier, R.D. 1996, MNRAS, 282,1299
\bibitem{}Tsiklauri, D., \& Viollier, R.D. 1998a, ApJ, 500, 591
\bibitem{}Tsiklauri, D., \& Viollier, R.D. 1998b, ApJ, 501, 486
\bibitem{}Tsiklauri, D., \& Viollier, R.D. 1999, Astropart. Phys., 12, 199
\bibitem{} Tupper, G.B., Lindebaum, R.J. \& Viollier, R.D. 2000, Mod. Phys.
Lett. A, 15, 1221
\bibitem{}Viollier, R.D. 1994, Prog. Part. Nucl. Phys., 32, 51
\bibitem{}Viollier, R.D., Leimgruber, F.R., \& Trautmann, D. 1992,Phys. Lett. B,
297, 132
\bibitem{}Viollier, R.D., Trautmann, D., \& Tupper, G.B., 1993, Phys. Lett. B,
306, 79
\bibitem{} Wardle, M., \& Yusef-Zadeh, F., 1992, ApJ, 387, L65
\bibitem{} Yusef-Zadeh, F., Melia, F. \& Wardle, M. 2000, Science, 287, 85
\end{thebibliography}
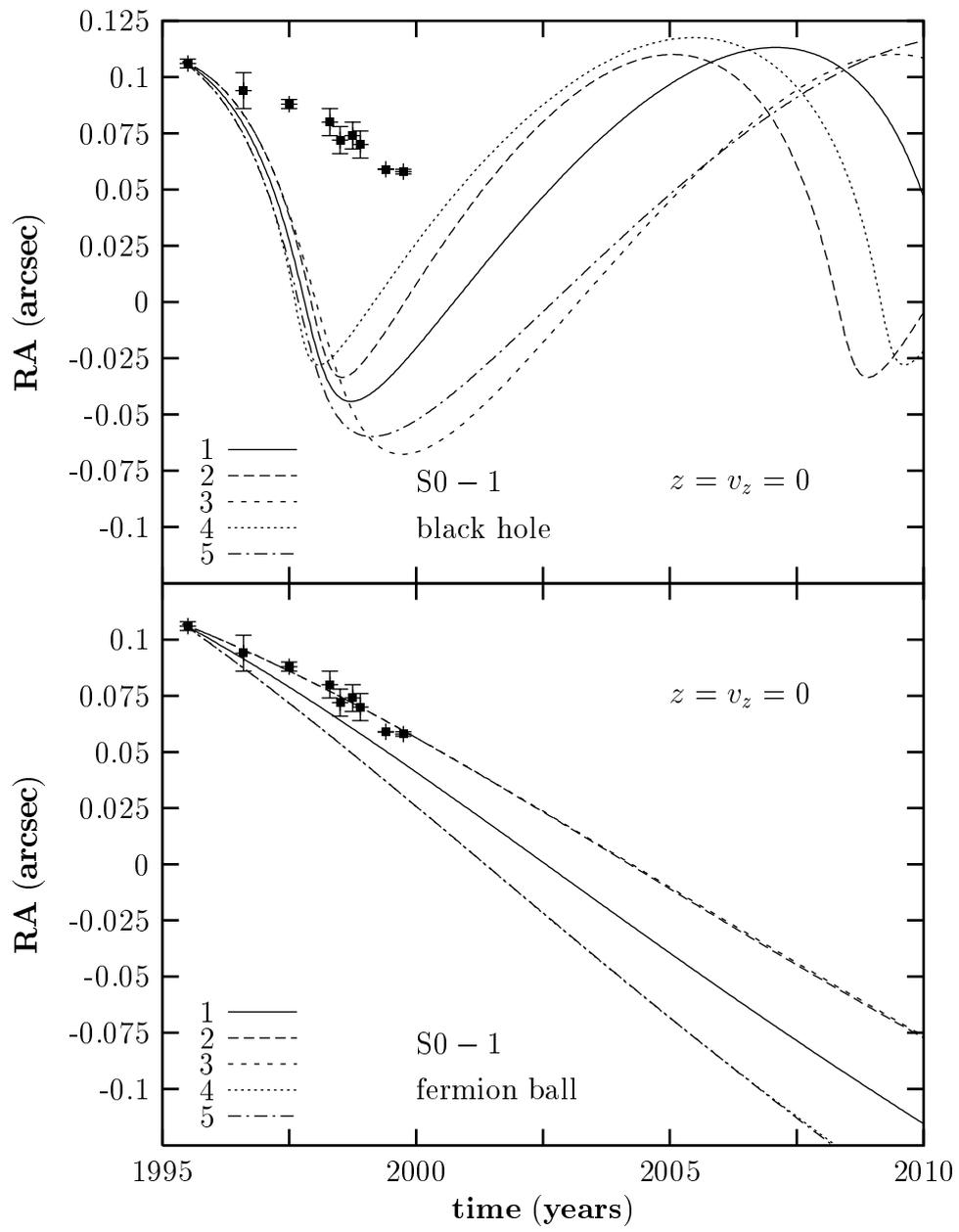
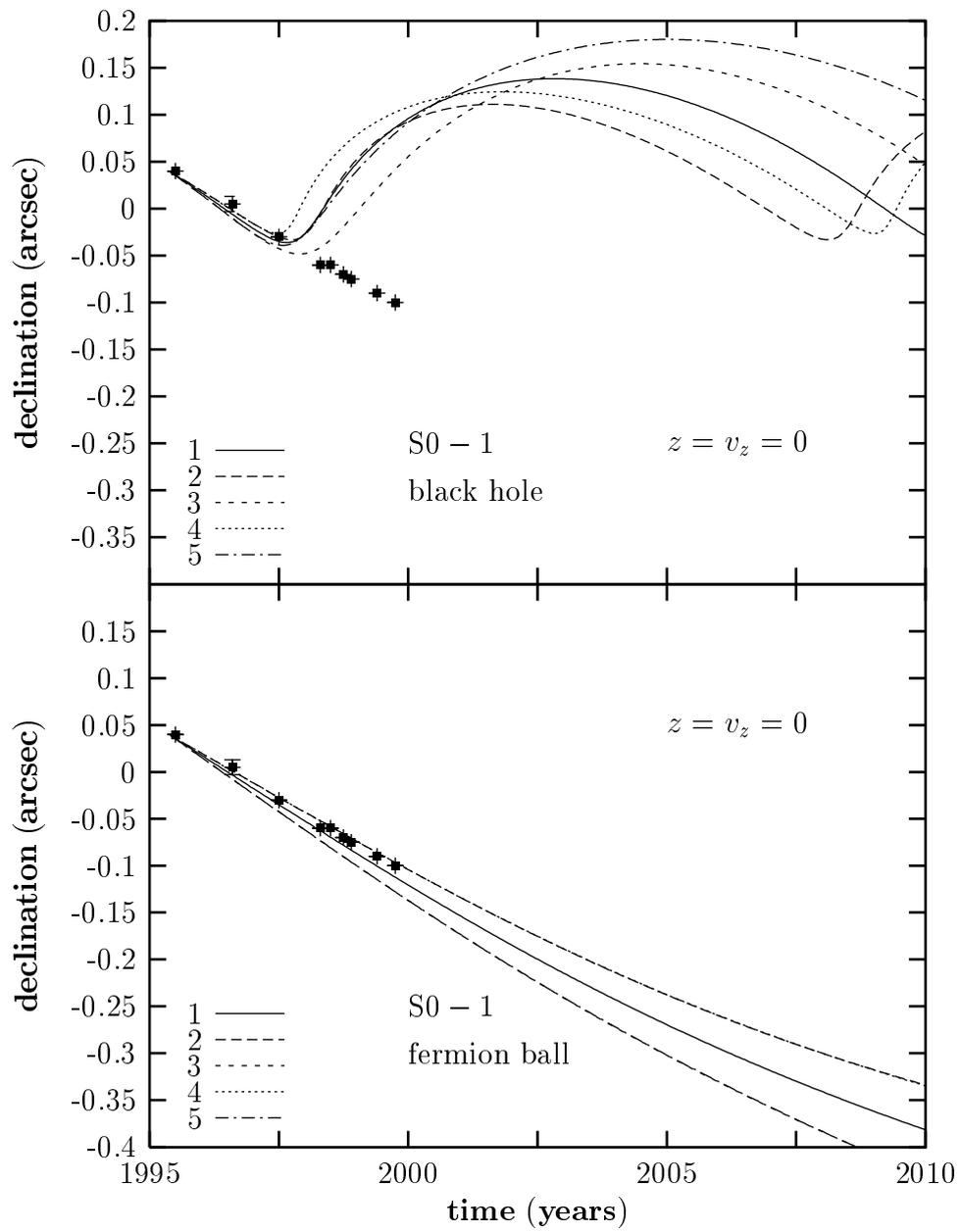
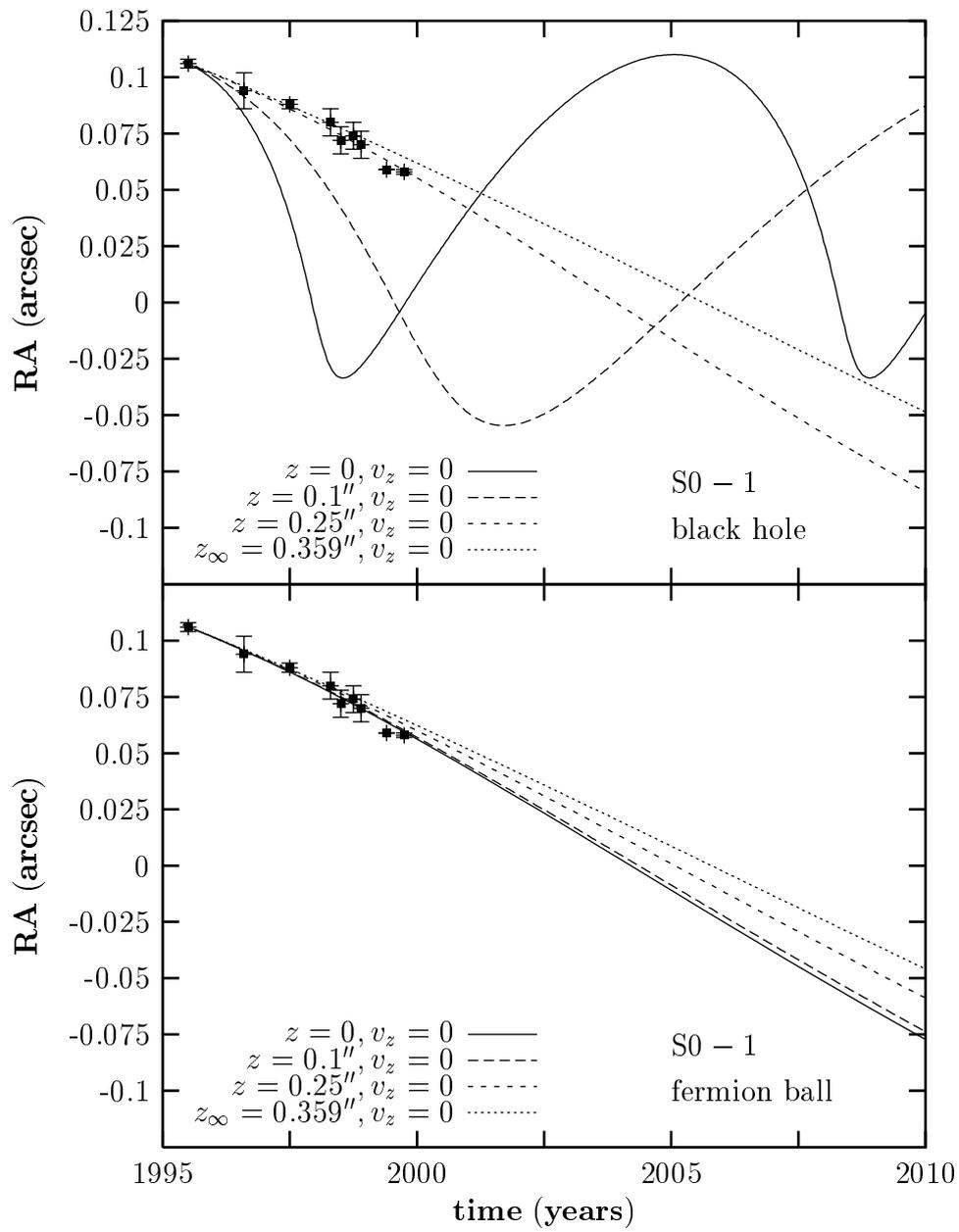
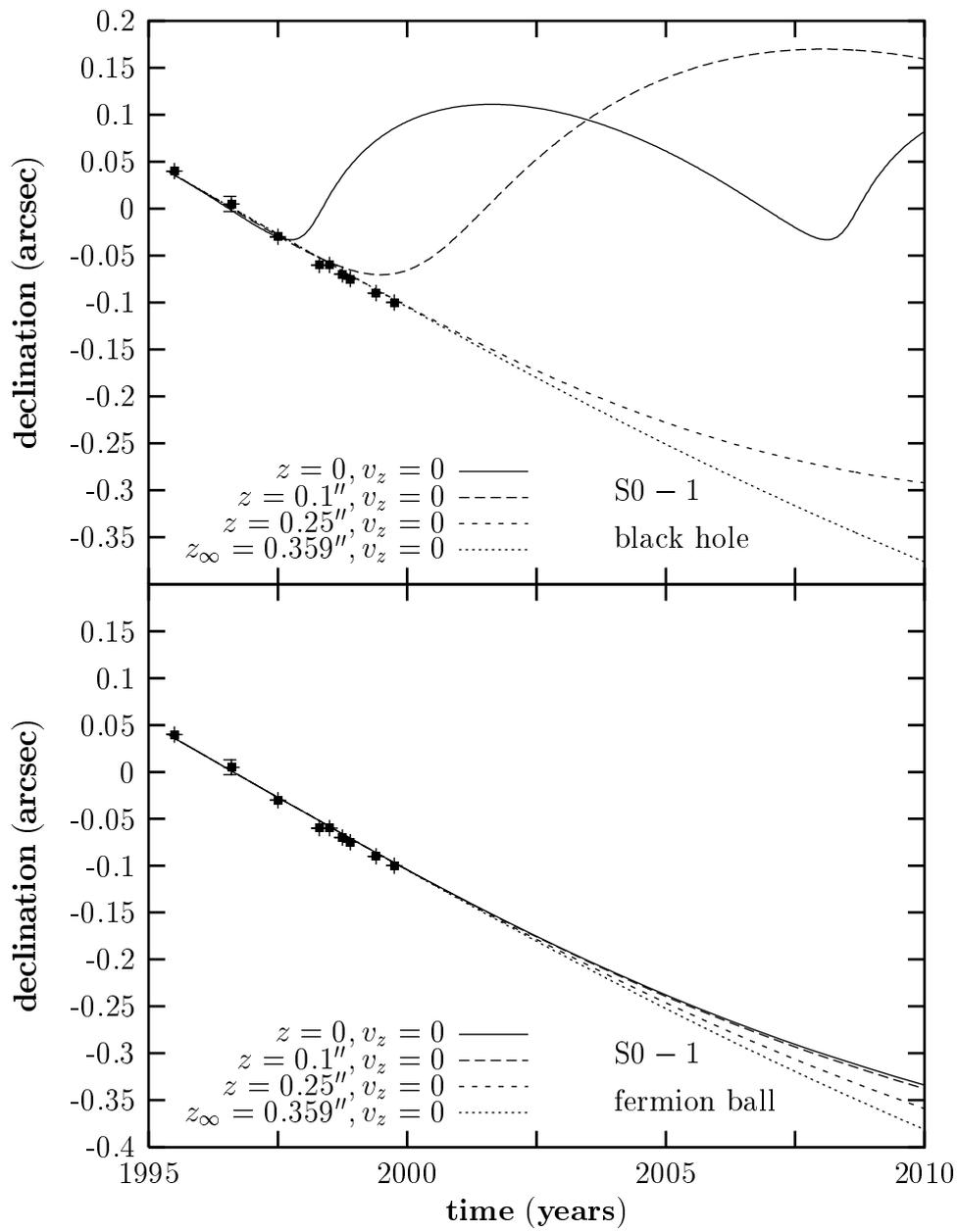
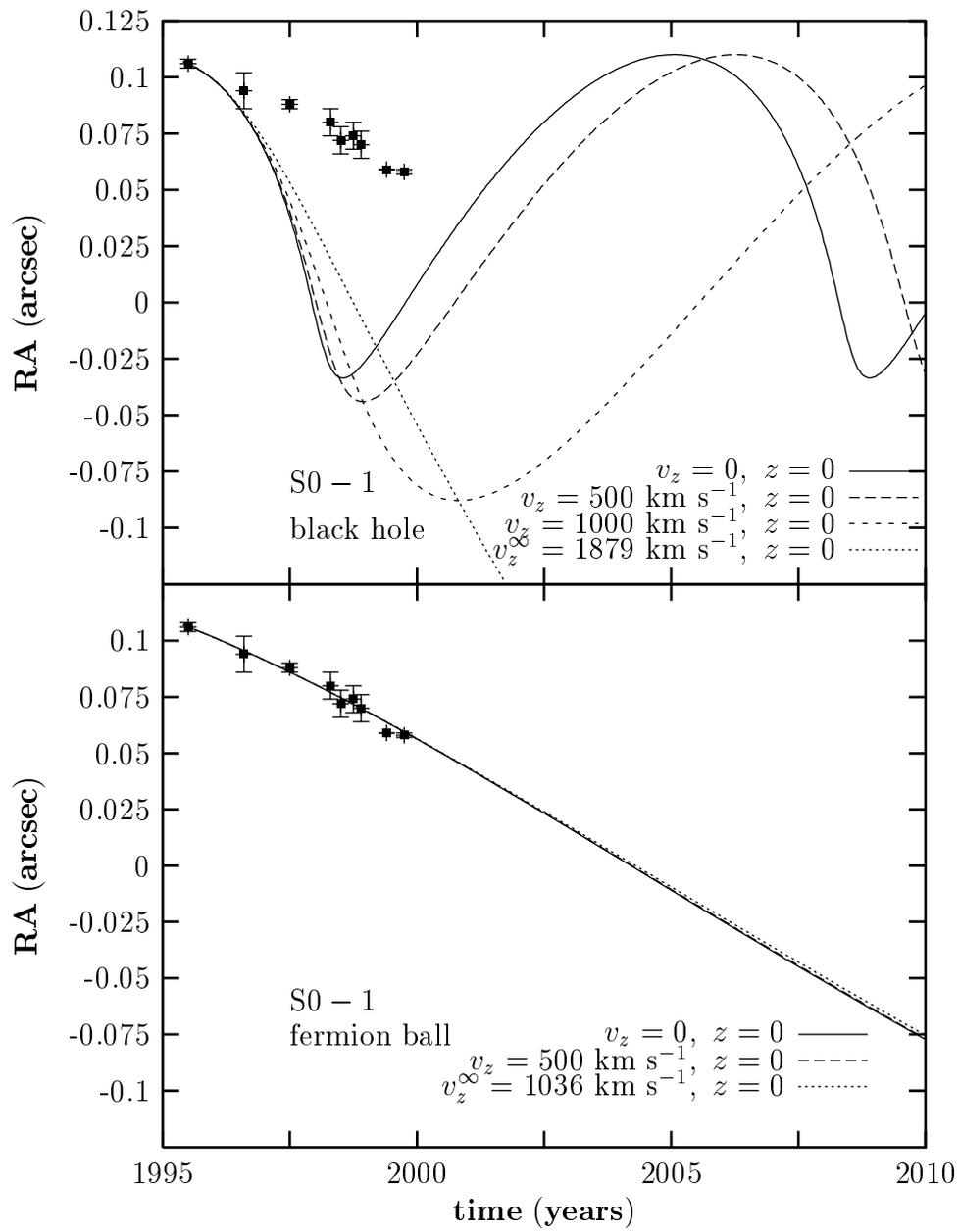
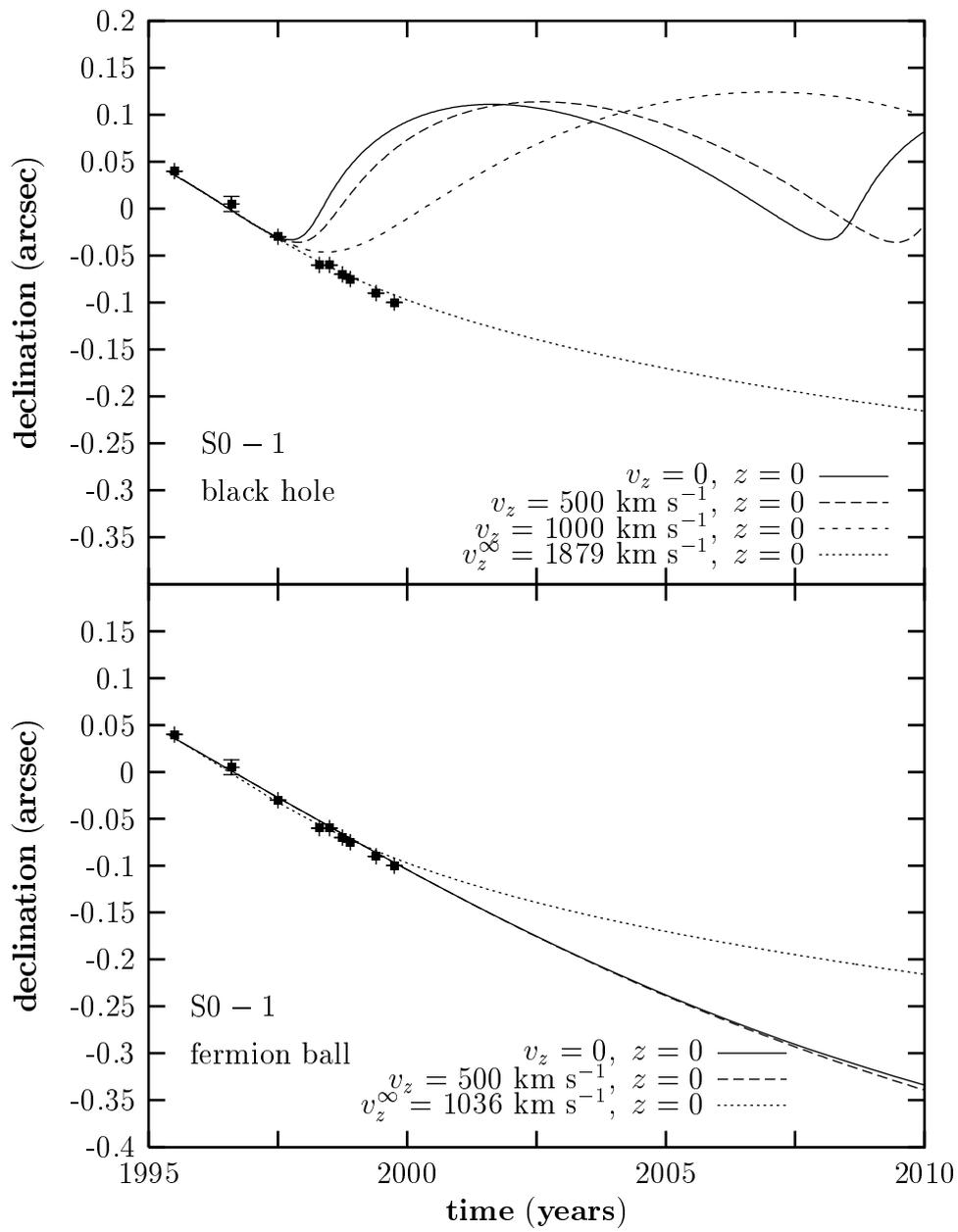
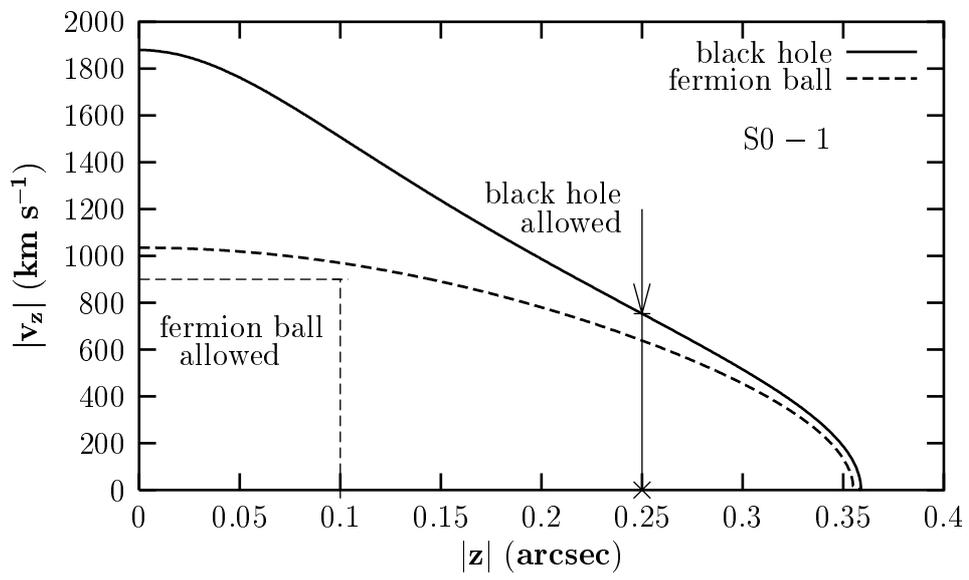
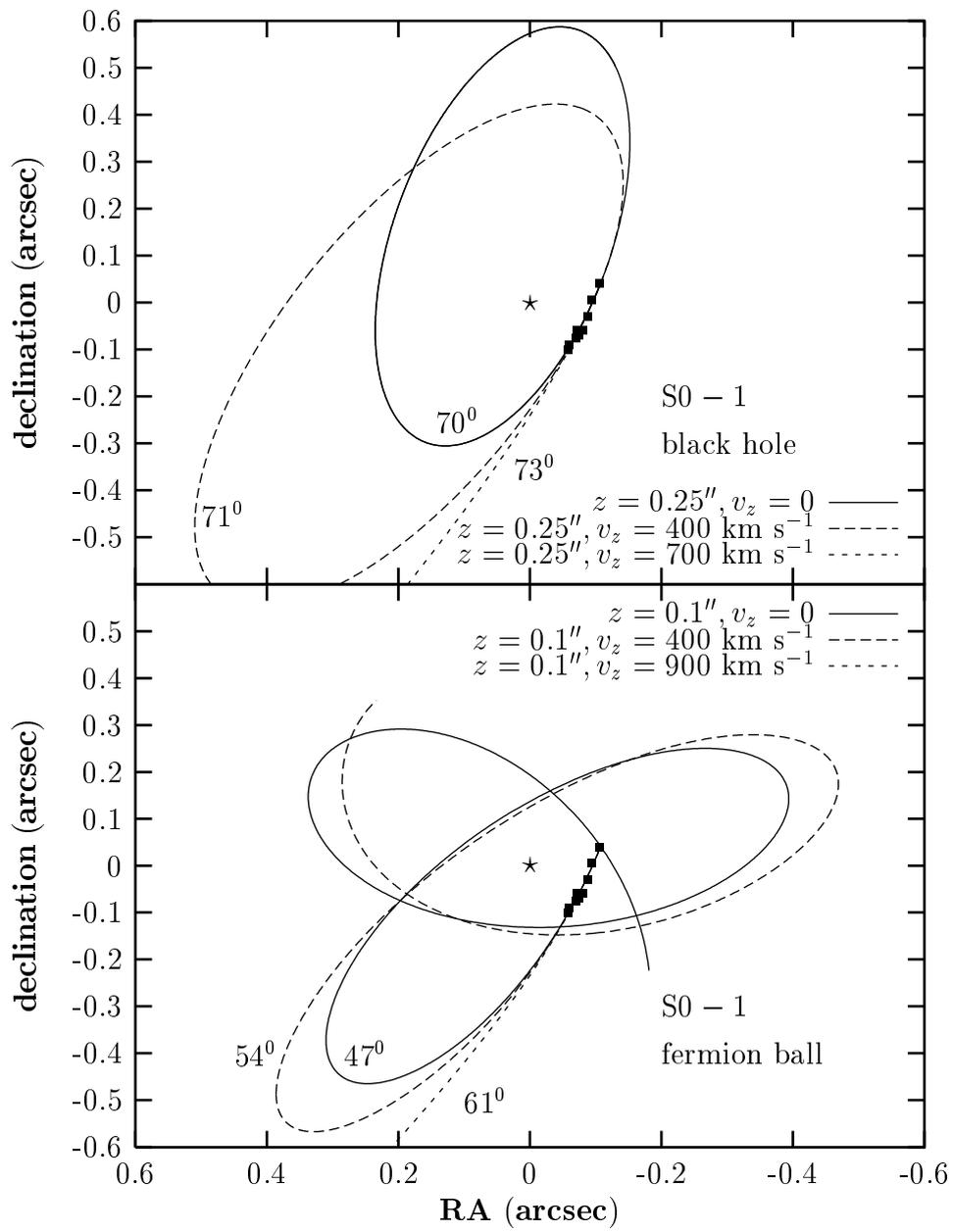
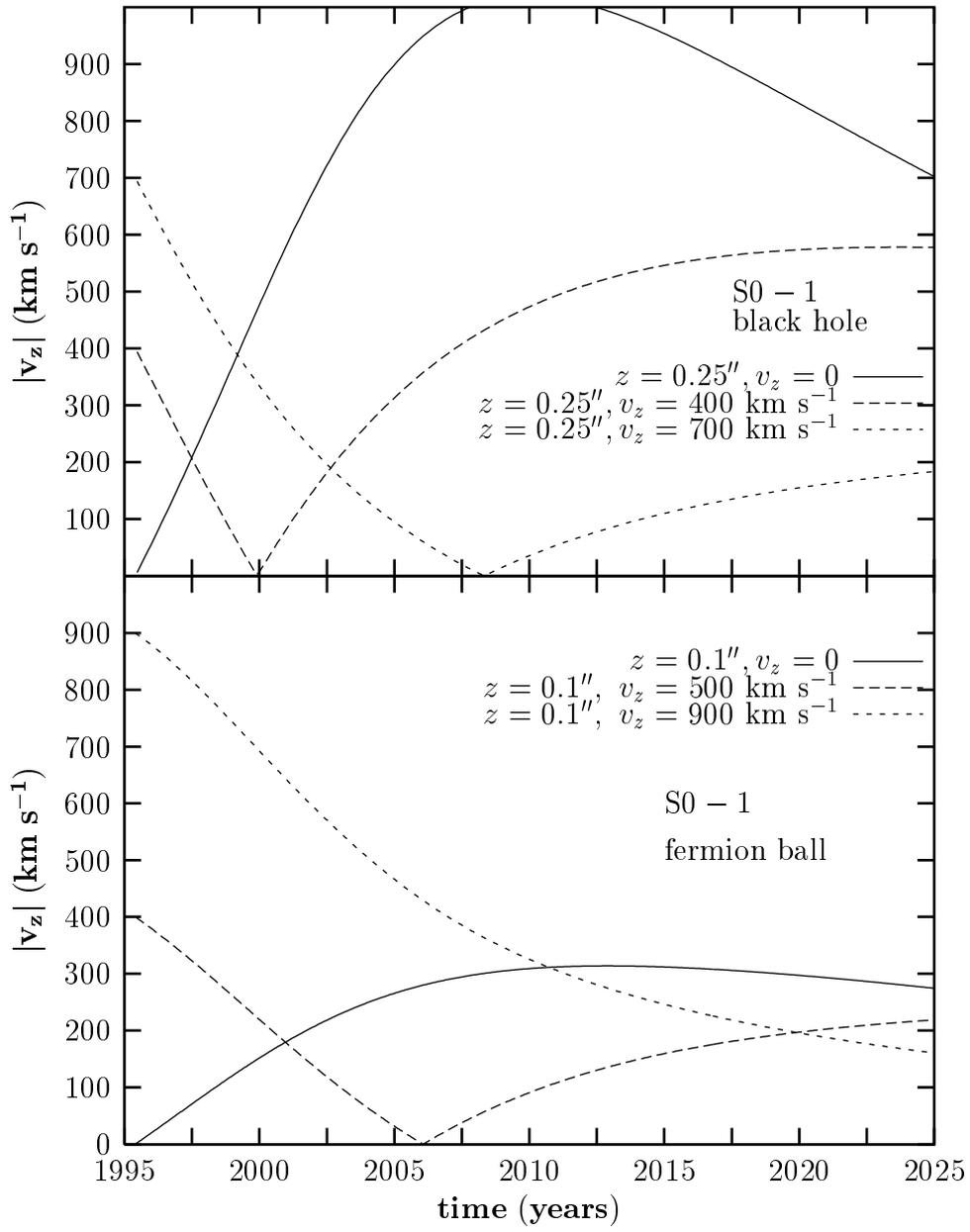
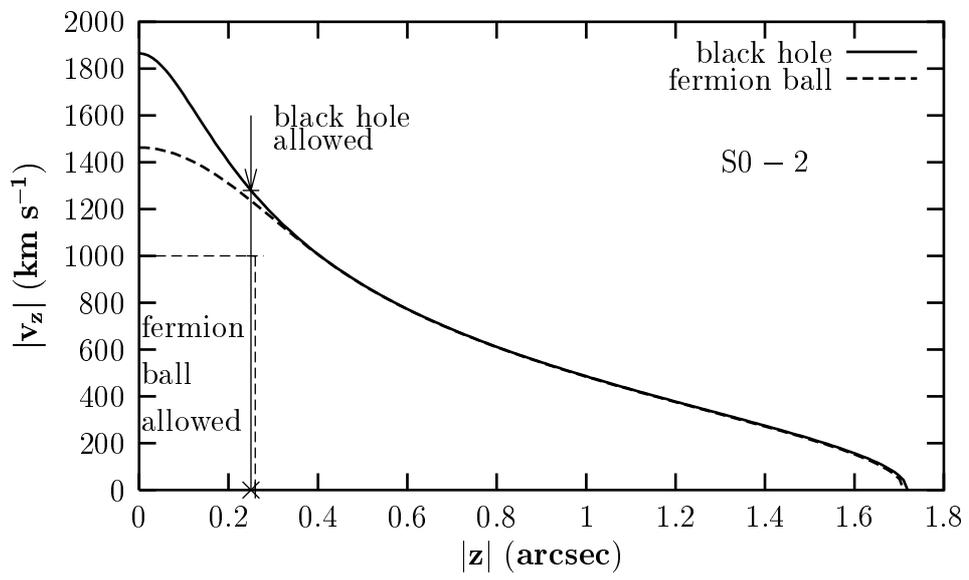
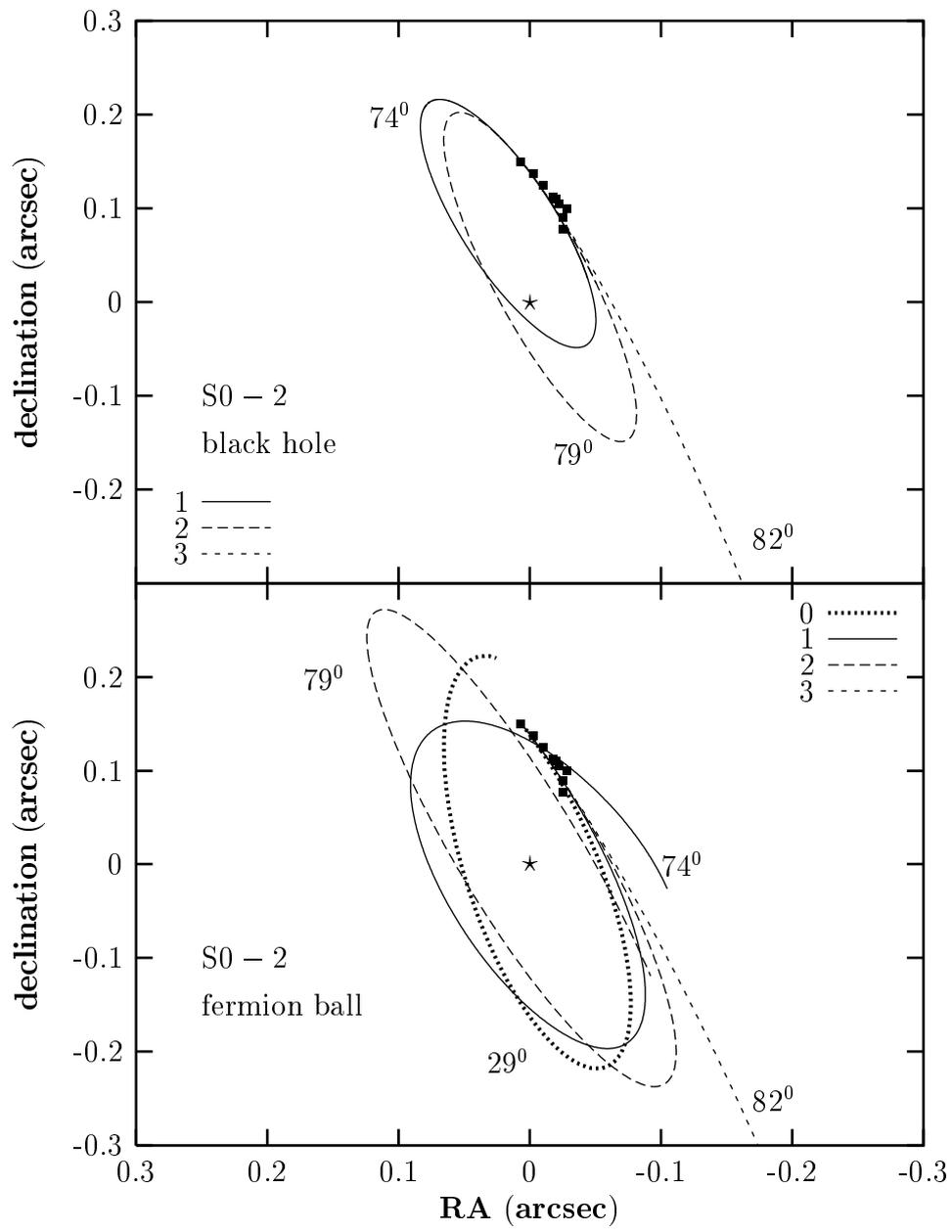
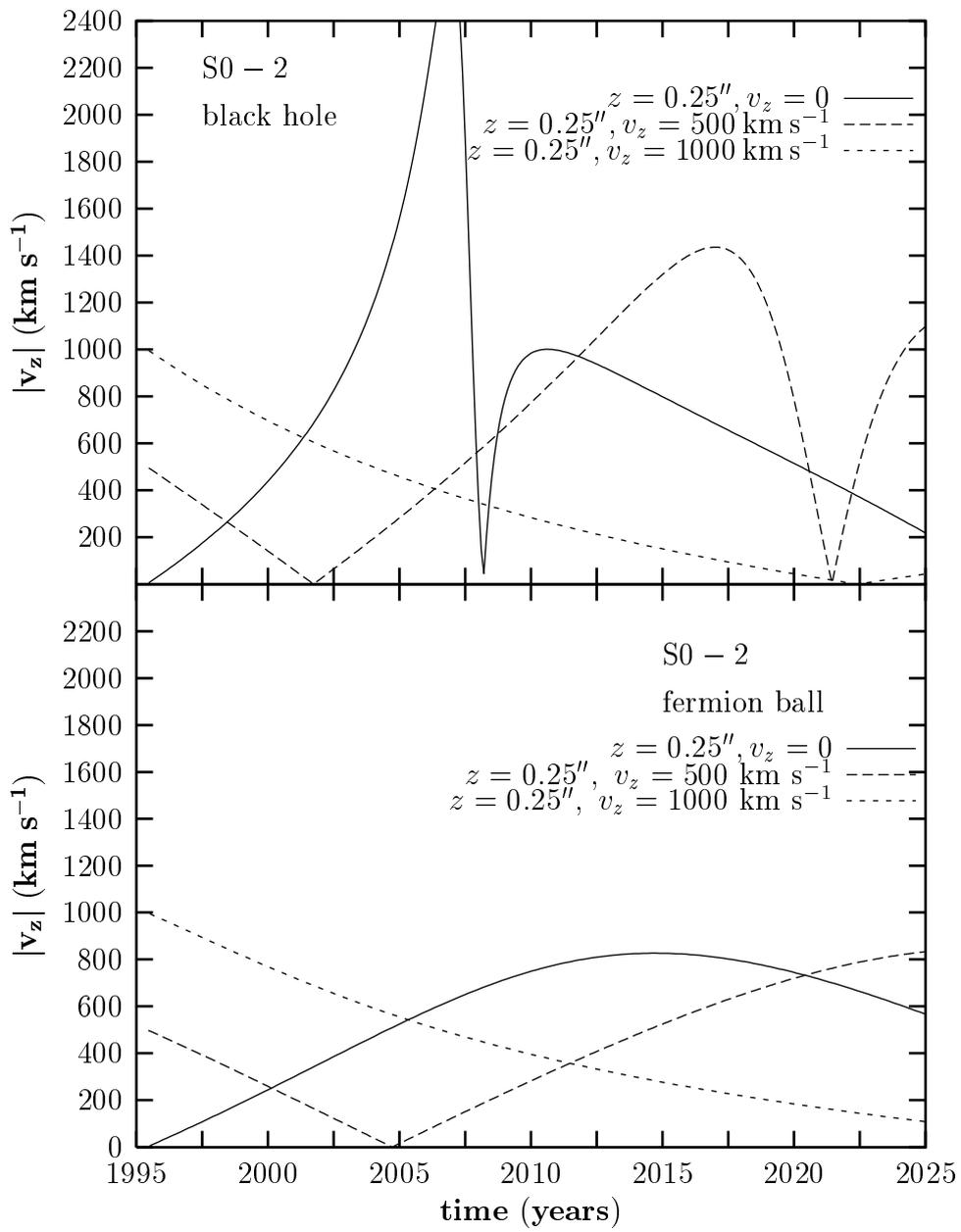
\end{document}